\documentclass[a4paper,11pt]{article}
\pdfoutput=1 

\usepackage{jheppub} 
\usepackage{lineno}
\usepackage{comment}
\usepackage{placeins}
\usepackage[all]{nowidow} 
\usepackage{xspace}

\usepackage{parskip}

\usepackage{siunitx}

\newcommand{\xlocal}{$x_{\mathrm{local}}$\xspace}
\newcommand{\ylocal}{$y_{\mathrm{local}}$\xspace}
\newcommand{\xsize}{$x_{\mathrm{size}}$\xspace}
\newcommand{\ysize}{$y_{\mathrm{size}}$\xspace}
\newcommand{\xprofile}{$x_{\mathrm{profile}}$\xspace}
\newcommand{\yprofile}{$y_{\mathrm{profile}}$\xspace}

\newcommand{\zglobal}{$z_{\mathrm{global}}$\xspace}

\newcommand{\nmodule}{$n_{\mathrm{module}}$\xspace}

\newcommand{\tcorr}{$t_{\mathrm{corr}}$\xspace}

\newcommand{\bibrej}[1]{$\mathcal{R}_{\mathrm{BIB,#1}}^{\mathrm{cluster}}$\xspace} 
\newcommand{\bibred}[1]{$\mathcal{R}_{\mathrm{BIB,#1}}^{\mathrm{data}}$\xspace}

\newcommand{\smartpixels}{\textit{smartpixels}\xspace}

\usepackage{graphicx}
\usepackage{subcaption}
\usepackage{pdfpages}
\usepackage{multirow}
\usepackage{enumitem}
\usepackage{siunitx}
\usepackage{changepage}
\usepackage{tabularx}
\usepackage{multirow}
\usepackage{array}

\renewcommand{\afterProceedingsSpace}{\vskip 0pt}

\title{On-Detector Machine Learning for Beam-Induced Background Rejection at a 10 TeV Muon Collider}


\affiliation[a]{The University of Chicago, Chicago, IL}
\affiliation[b]{Fermi National Accelerator Laboratory, Batavia, IL } 
\affiliation[c]{University of Illinois Chicago, Chicago, IL}
\affiliation[d]{Cornell University, Ithaca, NY}
\affiliation[e]{Purdue University, West Lafayette, IN}
\affiliation[f]{University of Illinois Urbana–Champaign, Champaign, IL}  
\affiliation[g]{Johns Hopkins University, Baltimore, MD}
\affiliation[h]{Northwestern University, Evanston, IL}  
\affiliation[i]{University of Colorado Boulder, Boulder CO}
\affiliation[j]{Northeastern University, Boston, MA}

\author[a]{Daniel Abadjiev,}
\author[a]{Eliza Howard,}
\author[a]{Tsz Ngong You,} 
\author[a]{Ryan Michaud,}
\author[a]{Benjamin Ryan Roberts,}
\author[a]{Benjamin Rosser,}
\author[a]{and Karri Folan Di Petrillo;}

\author[b]{Doug Berry,}
\author[c]{Arghya Ranjan Das,}
\author[d]{Jennet Dickinson,}
\author[b]{Giuseppe Di~Guglielmo,}
\author[c]{Harshul Gupta,}
\author[b]{Farah Fahim,}
\author[b]{Abhijith Gandrakota,}
\author[b]{Lindsey Gray,}
\author[b]{James Hirschauer,}
\author[f]{David Jiang,}
\author[e]{Shiqi Kuang,}
\author[b]{Ron Lipton,}
\author[a]{Mira Littmann,}
\author[e]{Miaoyuan Liu,}
\author[j]{Nicholas Manganelli,}
\author[g]{Petar Maksimovic,}
\author[c]{Corrinne Mills,}
\author[f]{Mark S.~Neubauer,}
\author[a]{Aidan Nicholas,}
\author[b]{Benjamin Parpillon,}
\author[i]{Jannicke Pearkes,}
\author[b]{Adam Quinn,}
\author[c]{Danush Shekar,}
\author[i]{Ricardo Silvestre,}
\author[b]{Chinar Syal,}
\author[g]{Morris Swartz,}
\author[b]{Nhan Tran,}
\author[c]{Amit Trivedi,}
\author[i]{Keith Ulmer,}
\author[c]{Mohammad Abrar Wadud,}
\author[d]{Benjamin Weiss,}

\emailAdd{karri@uchicago.edu}
\abstract{A 10~TeV Muon Collider is a compelling candidate for a future energy-frontier facility, offering unprecedented opportunities to explore the fundamental laws of particle physics. Muon decays in the collider ring produce intense beam-induced background (BIB) that can overwhelm detector occupancy and exceed readout bandwidth constraints. We investigate the potential of on-detector Machine Learning for BIB rejection in the vertex detector, exploiting pixel cluster shapes to distinguish background from collision products. We study three classes of lightweight neural-network architectures, and evaluate their implementation feasibility using high-level synthesis. Selected architectures achieve 88--90\% data reduction at 99\% signal efficiency, while requiring hardware resources compatible with potential ASIC implementation. These results demonstrate the potential of performing substantial BIB rejection directly in the pixel readout, providing a strategy for meeting the tracker readout requirements at a future Muon Collider.}


\begin{document}
\maketitle
\flushbottom

\section{Introduction}
\label{sec:intro}
A 10 TeV Muon Collider is a compelling option for the future of the energy frontier. Muon Colliders combine precision and energy reach in a single compact and power-efficient machine. A Muon Collider would also deepen our understanding of the Higgs boson, electroweak symmetry breaking, and minimal dark matter models~\cite{TowardsMuC, themuoncollider}. However, detector occupancy due to beam-induced background (BIB) is anticipated to exceed readout constraints, in particular for the pixel detector. This paper investigates the use of Machine Learning (ML) implemented in custom readout electronics to achieve real-time on-detector data reduction, extending the approach of the \smartpixels~\cite{filtering,testing,codesignPublication} collaboration beyond the High Luminosity Large Hadron Collider (HL-LHC) environment. 

The primary challenge for a Muon Collider experiment is posed by muon decays in the beam upstream of the detector. Tungsten nozzles in the forward region shield the detector from the resulting multi-TeV electrons but produce showers of low-energy neutrons, photons, and electrons that leak into the detector volume each bunch crossing. On average, $\mathcal{O}(10^8)$ BIB particles are anticipated per event, posing significant challenges for detector read-out, reconstruction, and radiation hardness~\cite{TowardsMuC, themuoncollider, maia, music}.

In the pixel detector, the majority of hits are produced by low energy electrons and positrons that propagate from the nozzles through the detector volume. With a mean momentum of $\mathcal{O}(10)~\mathrm{MeV}$, 
a single particle's helical trajectory can produce multiple hits within an individual layer of the tracker. The exact occupancy due to BIB will depend on the interaction region optics, tungsten nozzles, and tracker geometry. Current 10 TeV designs anticipate $\mathcal{O}(10^4)$~\unit{hits/cm^2} in the innermost layer of the vertex detector per bunch crossing~\cite{themuoncollider}. Assuming a 30 kHz bunch crossing frequency, front-end chip area $2\times2~\mathrm{cm}^2$, and 40-bit hit data size, this occupancy translates to data rates of over ${\sim}50~\mathrm{Gbps}$ per chip, representing a factor of ten increase over the maximum readout rate per chip for High Luminosity LHC pixel detectors~\cite{TowardsMuC, itkPixTDR}. While advancements in telecommunications technology could enable higher readout rates, a readout suitable for the Muon Collider pixel detector which meets material, power, and radiation constraints has not been demonstrated.



Maximizing discovery potential calls for triggerless readout with real-time on-detector data reduction~\cite{promisingtechnologiesrddirections}. 
Several studies have demonstrated that pixel detector occupancy can be reduced to manageable levels for reconstruction by requiring hits to arrive in time with respect to particles produced in the collision. However, these studies assume $\mathcal{O}(30)~\unit{ps}$ precision timing resolution. While this level of precision has been demonstrated for $1\times1~\mathrm{mm}^2$ pixels,  $25\times \SI{25}{\micro\meter}^2$ pixels would require at least an order-of-magnitude reduction in front-end power consumption per channel~\cite{Affolder:2022qll}. Moreover, using timing as the primary means of BIB rejection assumes that sufficiently precise and stable timing information can be maintained. Achieving this online will be challenging because the timing measurement is affected by clock synchronization, electronics stability, radiation damage, calibration drift, and high occupancy, motivating alternative approaches to on-detector BIB rejection. 

This paper aims to demonstrate that we can use differences in pixel cluster shapes to reduce data rates to manageable levels with negligible loss of hits from collision particles. Particles produced in collisions tend to traverse the detector with incident angles consistent with the interaction point. In contrast, BIB particles tend to traverse pixel layers with large incident angles regardless of hit location. Because the shape and size of a pixel cluster depend strongly on the particle's incident angle, the resulting clusters can be exploited for classification. In addition to meeting requirements on signal efficiency and background rejection, a realistic algorithm must be implementable in the readout hardware within the power, latency, material, and other constraints of the detector. These objectives are in tension, as more complex methods of BIB rejection generally require more resources. We investigate the performance and hardware usage of a range of neural network algorithms for this classification task. 


The \smartpixels  collaboration has previously investigated ML-based on-detector data reduction, focused on a hadron collider environment. Studies in simulation have demonstrated that neural networks trained on pixel cluster shapes can be used to either filter hits produced by low momentum particles or regress an incident particle's trajectory ~\cite{filtering,regression}. In addition, the \smartpixels concept has been demonstrated in a prototype application specific integrated circuit (ASIC) with analog signal processing and a digital neural network ~\cite{smartpixelprototype}. Complementary studies at a Muon Collider have used fixed selections on pixel hit time of arrival or cluster size to reject clusters produced by BIB~\cite{angiraCHECKCITATION}.
Here, we investigate using neural networks to filter clusters produced by BIB in the innermost tracking layer of a Muon Collider detector.  

\section{Simulated Datasets}
\label{sec:datasets}


ML algorithms are trained and evaluated on two independent, labeled samples of simulated pixel clusters. The signal sample consists of single muons originating at the interaction point. The background sample consists of BIB particles generated externally by the International Muon Collider Collaboration (IMCC). Signal and BIB particles are both propagated through the same general-purpose detector geometry with \texttt{GEANT4} (via \texttt{DDSim})~\cite{geant4,dd4hep}. Simulated hits are converted into pixel charge clusters with \texttt{PIXELAV}~\cite{pixelavNote}. Section~\ref{sec:datasetFeatures} defines the per-cluster features derived from this pipeline that serve as inputs to ML filtering algorithms.



Signal particle kinematics are sampled from uniform distributions in momentum $p$, polar angle $\theta$, and azimuthal angle $\phi$. Momenta considered range from 1 GeV, which represents a reasonable minimum target for track reconstruction, to 100 GeV, which corresponds to an approximately straight track in the first pixel layer. The $\theta$ range is restricted to that of the innermost barrel layer of the pixel detector, \ang{25} to \ang{155}. To simulate a realistic beam spot size, particle production points are sampled from Gaussian distributions centered at the origin of the detector with a spread of $\sigma_z = 1.5~\mathrm{mm}$ along the beamline and $\sigma_{xy}=0.9$ \unit{\micro\meter} in the transverse plane~\cite{TowardsMuC}. 

BIB particles and trajectories are provided by the IMCC. The simulation is performed from the beam line to the beginning of the detector region with \texttt{FLUKA} line builder~\cite{fluka1,fluka2,flukaLineBuilder}, using a collider lattice designed for $\sqrt{s} = 10~\mathrm{TeV}$ collisions, tagged version 0.4~\cite{Skoufaris:2022wwg}. While updates to the collider lattice and machine detector interface may influence the overall rate of BIB particles produced, they are not expected to significantly change distinguishing features. 

\begin{figure}
    \centering
    \includegraphics[width=0.94\linewidth]{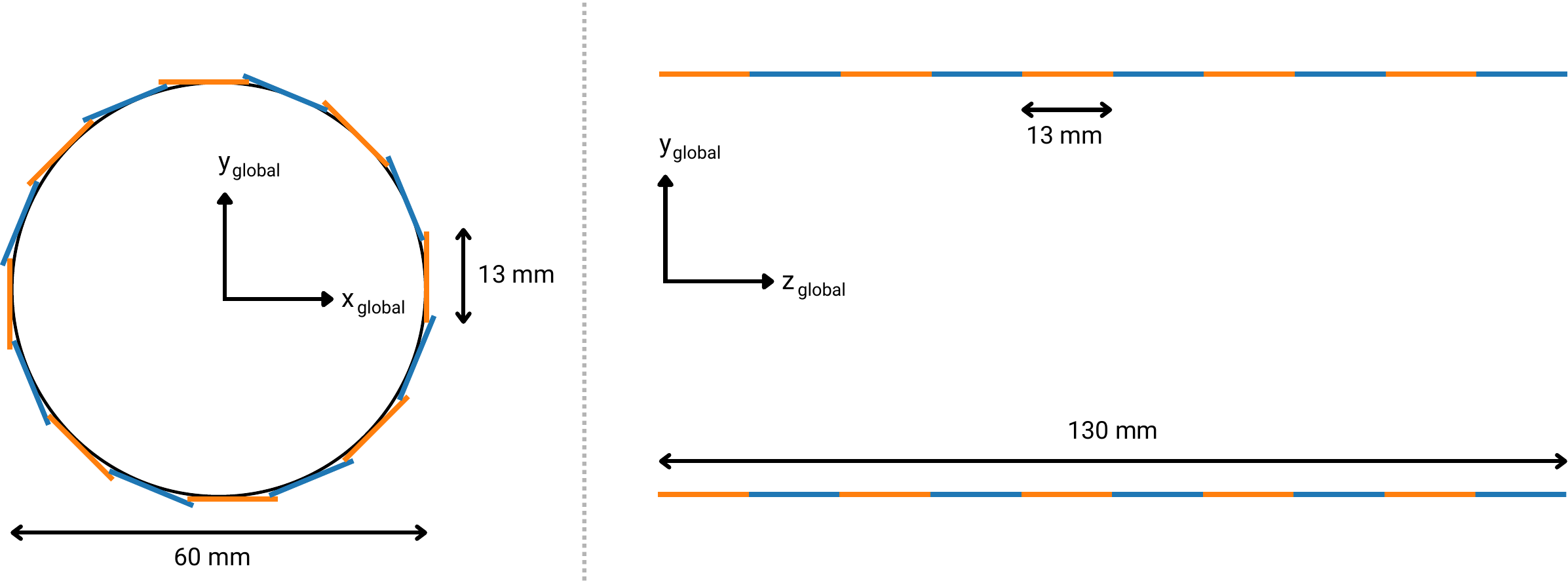}
    \caption{Pixel module arrangement for the first layer of the barrel Vertex Detector in the global $x$-$y$ (left) and $y$-$z$ (right) planes. Alternating colors distinguish adjacent modules.}
    \label{fig:sensorgeometry}
\end{figure}

The passage of signal and BIB particles through the detector is simulated with \texttt{GEANT4} using \texttt{DDSim}~\cite{dd4hep,geant4}. The \texttt{MuColl\_v1} geometry and a 3.57 T solenoidal magnetic field are assumed~\cite{TowardsMuC}. At this stage, Bremsstrahlung electrons are removed from the signal sample in order to ensure the remaining particles consist only of muons originating from the interaction point. For this initial study, all particles are also required to pass through the first layer of the vertex barrel detector, which experiences the highest rate of BIB. Modules are assumed to be $13 \times\SI{13}{\milli\meter}^2$ with pixel sizes of $25\times\SI{25}{\micro\meter}^2$ and $\SI{50}{\micro\meter}$ thick. Modules are positioned at a radius of $\SI{30}{\milli\meter}$ and offset in $\phi$ by an angle of \ang{11}. The layer extends to $|z|=\SI{65}{\milli\meter}$. The arrangement of modules is visualized in Figure~\ref{fig:sensorgeometry}. 

\begin{figure}
        \centering
        \includegraphics[width=0.94\linewidth]{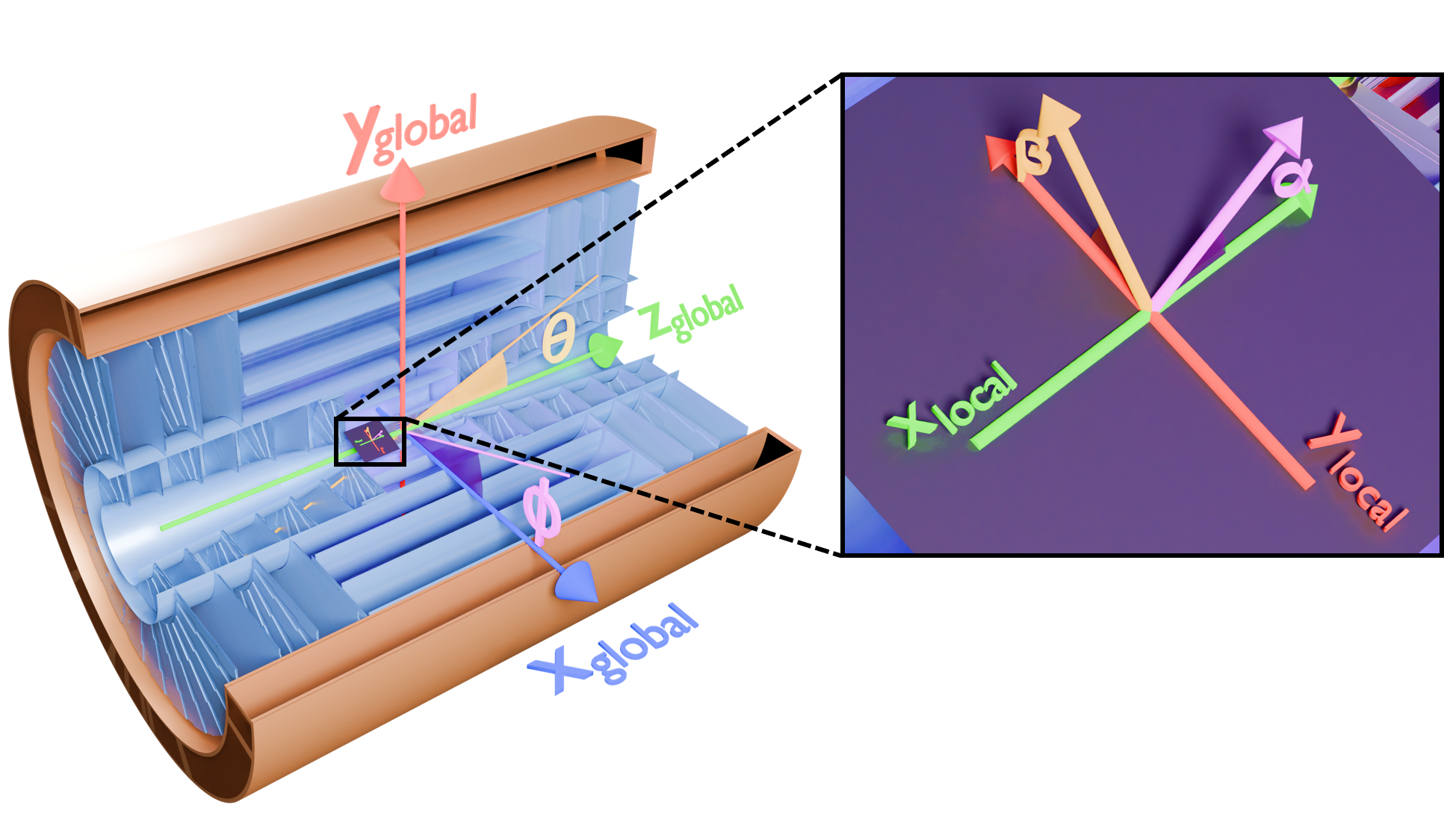}
        \caption{Diagram of coordinates in the detector barrel frame and their correspondence with local module coordinates.}
        \label{fig:coordinates}
\end{figure}

Each simulated hit and incident angle is then converted into the local coordinate system of the respective module. The correspondence between global and local coordinate systems is shown in Figure~\ref{fig:coordinates}.
Local $\hat{x}$ is parallel to global $\hat{z}$, local $\hat{y}$ is parallel to global $\hat{\phi}$, and local $\hat{z}$ is perpendicular to the module surface, parallel to the global cylindrical radius $\hat{r}$.
Simulated hits were also required to be produced by particles with both transverse and total momenta greater than $50~\mathrm{keV}$ and with incidence angles above \ang{5} to remove particles which would not produce a signal above threshold and excessively long clusters, respectively. 

The pixel detector's response to incident charged particles is simulated with \texttt{PIXELAV}~\cite{pixelavNote}. \texttt{PIXELAV} provides a detailed simulation of hybrid pixel detector response by incorporating accurate charge deposition from minimum ionizing particles, realistic semiconductor electric field maps, charge transport physics, and the resulting electronic noise, signal response, and threshold behavior. 
Hits which produce clusters with more than 150,000 electron hole pairs were removed due to computational constraints of \texttt{PIXELAV}. 

The simulation is restricted to a 21 \texttimes \ 13 array of pixels, referred to as the region of interest (ROI), and assumes the particle crosses the center of the module in the local $z$ direction within a 3 \texttimes \ 3 pixel region in the center of the ROI. The final output of \texttt{PIXELAV} is a 21 \texttimes \ 13 array of the total number of electron hole pairs collected in each pixel as a function of time. Following simulation, the ROI location is re-positioned to be centered at the charge centroid, such that the location is defined by detector-level information. The simulated datasets produced for this study are available in \cite{Smartpix4MuCZenodo}.


Several simplifying assumptions are made for this initial study. The simulation by \texttt{PIXELAV} assumes a clean ROI which only contains a single particle hit throughout a 4 \unit{ns} front-end response time. However, the  anticipated occupancy due to BIB could result in nearby or overlapping clusters. We do not overlay signal and BIB clusters, nor do we consider effects of clusters warped by charge deposits from more than one hit, which could occur. Finally, we assume that the ROI is always well-defined with no module edge effects, and that previous logic locates the appropriate ROI for each hit.


In this study, hits are required to have a corrected time of arrival within $[-0.5 \ \unit{ns}, 15 \ \unit{ns}]$. The corrected hit time is defined as $t_{\mathrm{corr}}= t_\mathrm{hit} - L/c$, where $t_{\mathrm{hit}}$ is the measured hit time relative to the collision, $L$ is the distance from the interaction point to the hit, and $c$ is the speed of light. For relativistic particles produced promptly at the interaction point, \tcorr should be centered at zero, as is the case for signal particles. Prior to this selection, hit times are smeared with a 30 ps gaussian to simulate the detector resolution. This smearing and baseline timing selection is consistent with previous Muon Collider detector studies~\cite{maia}.



\begin{figure}
    \centering
    \includegraphics[width=0.75\linewidth]{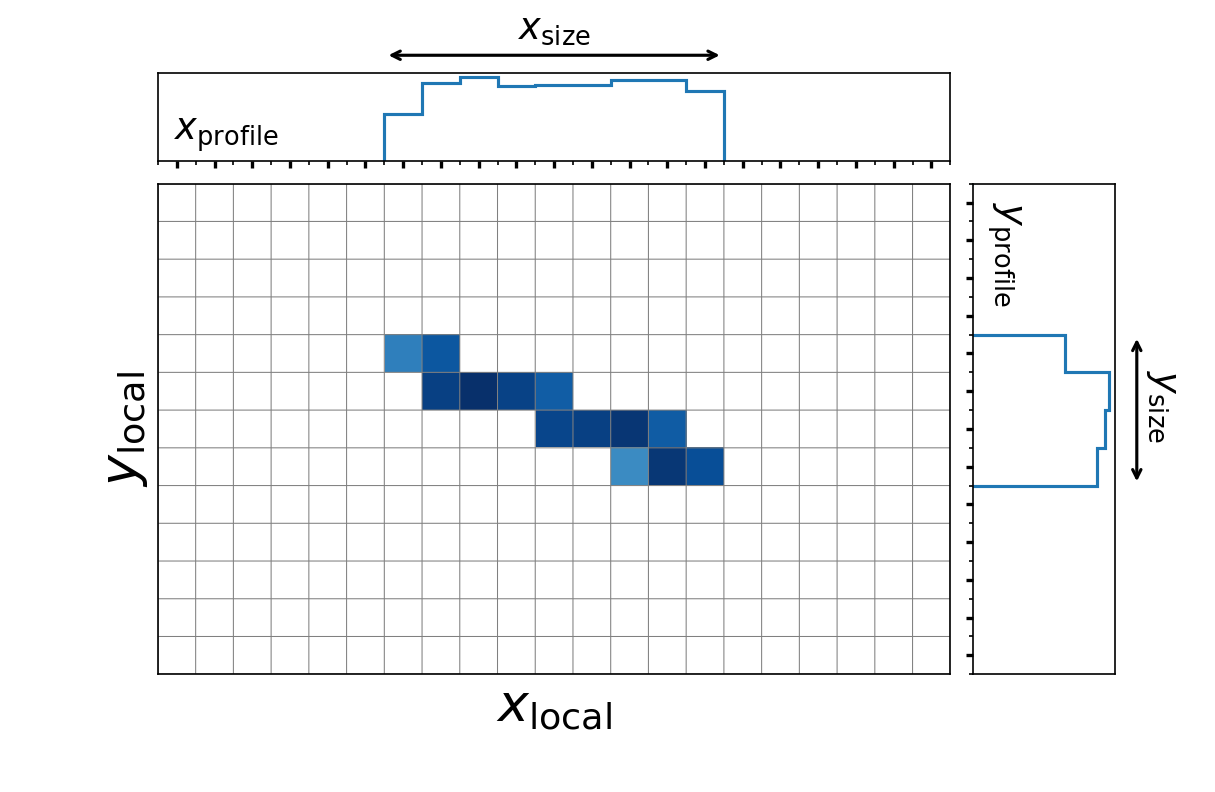}
    \caption{Schematic of a BIB cluster simulated by \texttt{PIXELAV}. The ROI, along with local coordinates and engineered cluster features described in \ref{sec:datasetFeatures}.}
    \label{fig:clusterFeatures}
\end{figure}

\subsection{Data Set Features}\label{sec:datasetFeatures}

Clusters from \texttt{PIXELAV} and their corresponding position information are converted into Tensorflow records for ML training and evaluation. For both BIB and signal, 1,656,656 clusters were produced. To facilitate training, clusters were shuffled and inputs were normalized. The dataset was divided into 80\% training and 20\% validation samples.

For each signal or BIB cluster we consider the following input features:
\begin{itemize}[itemsep=0em, parsep=1pt] \vspace{-0.5em}
    \item \textit{charge cluster}: A $21 \times 13$ array containing the charge collected by each pixel of the ROI, in units of electrons.
    
    \item \zglobal: The global $z$-coordinate of the ROI in the overall detector coordinate system. Since this coordinate has a high dynamic range, it can be further subdivided into a (coarse) module number and (fine) local coordinate. 
    \item \nmodule: The module number in the longitudinal direction, which is parallel to \zglobal.
    \item \xlocal: The local $x$-coordinate of the ROI in the module reference frame, which is parallel to \zglobal.
    \item \ylocal: The local $y$-coordinate of the ROI in the module reference frame.

\end{itemize}

We also define a set of \emph{engineered features} that summarize cluster shapes in one dimension, visualized in Figure~\ref{fig:clusterFeatures}:
\begin{itemize}[itemsep=0em, parsep=1pt] \vspace{-0.5em}
    \item \xprofile: A one-dimensional histogram of the cluster in the \xlocal direction, obtained by summing charge collected in each column of pixels. 
    \item \yprofile: A one-dimensional histogram of the cluster in the \ylocal direction, obtained by summing the charge collected in each row of pixels. 
    \item \xsize: A compact measure of the cluster width in the \xlocal direction, defined as the number of $x$ bins in the \xprofile with non-zero charge. 
    \item \ysize: A compact measure of the cluster width in the \ylocal direction, defined as the number of $y$ bins in the \yprofile with non-zero charge. 
\end{itemize}

\begin{figure}
        \centering
        \includegraphics[width=0.95\linewidth]{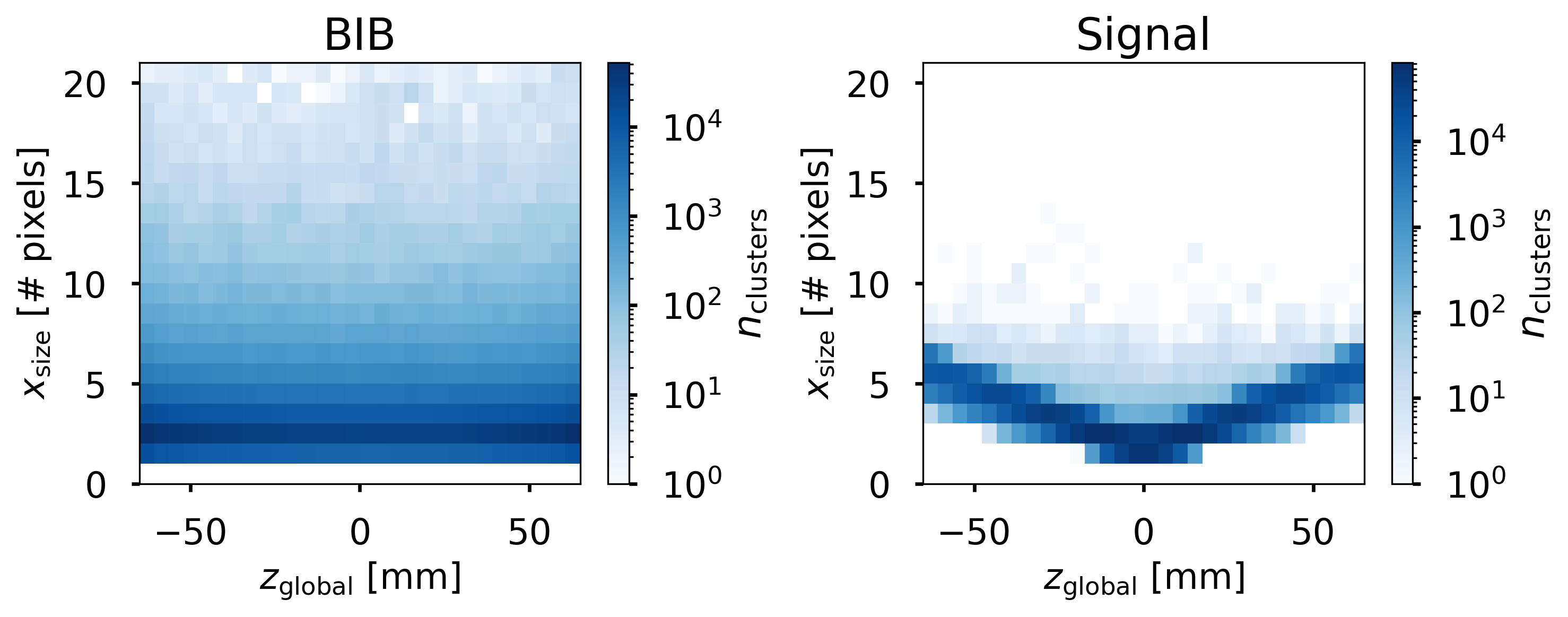}
        \caption{Distributions of BIB (left) and signal (right) cluster \xsize as a function of \zglobal. }
        \label{fig:xsizezglobal}
\end{figure}


Figure~\ref {fig:xsizezglobal} shows key features of signal and background clusters. Signal clusters have incident angles consistent with production at the interaction point. In particular, the \xsize of the cluster increases with \zglobal. In contrast, BIB particles looping in from the sides of the detector produce clusters with large \xsize and \ysize regardless of the detector location. Additional features of the BIB and signal datasets are shown in Appendix~\ref{app:dataset}. 

\section{BIB Filtering Algorithms}

In order to enable on-detector data reduction, we develop three classes of neural network 
architectures of increasing complexity 
trained to classify clusters as BIB or signal using the geometric features described in Section~\ref{sec:datasetFeatures}. 
Multiple architectures were investigated for their potential to be implemented in digital logic, consistent with a low-power, size-constrained ASIC. In particular, we explore the trade-offs between classification performance and hardware resource usage. 

The rate of background clusters rejected for fixed signal cluster efficiencies of 95\%, 98\%, and 99\% is used to evaluate model performance. Achieving the excellent tracking performance called for in preliminary detector designs will require the readout chip to achieve very high signal efficiency, as is the case for the High Luminosity LHC pixel detectors \cite{maia,music,itkPixTDR}. Therefore, background rejection at 99\% signal efficiency, denoted \bibrej{99}, is used as the metric for hyperparameter optimization and the primary metric for evaluating model performance. ML performance is also compared to fixed selections on hit time of arrival and cluster size.



\subsection{Time of Arrival Selection}

Unlike signal, most BIB particles produce hits out of time with respect to the collision, and precision timing has been demonstrated as a powerful handle for rejecting BIB.
Several previous studies have assumed a pixel detector timing resolution of $\sigma=30\ \unit{ps}$, and apply selections on the corrected hit time to reject clusters produced by BIB. Requiring corrected hit times within $[-3\sigma,5\sigma]$ has been shown to reduce occupancy due to BIB by an order of magnitude \cite{maia}. We refer to this selection as the \texttt{nominal} timing window.

For a baseline comparison to previous work, we investigate the impact of applying the  \texttt{nominal} timing window to our samples. This selection achieves a signal efficiency of 99.9\% and a background rejection of 88.0\%, consistent with previous studies \cite{angiraCHECKCITATION}.
To more fully characterize timing-based selections, we also consider requirements of $[-3\sigma, T]$, varying the upper bound $T$ to investigate BIB rejection at fixed signal efficiency. Results are summarized in Table~\ref{table:timing}, highlighting background rejection at 95\%, 98\%, and 99\% signal efficiency to facilitate direct comparison with neural network performance metrics. The asymmetric window is chosen for consistency with previous Muon Collider studies~\cite{maia, angiraCHECKCITATION}.




\begin{table}[h]
\centering
\begin{tabular}{c|c|c}
\hline
\hline
 \textbf{Timing Selection} & \textbf{Signal Efficiency} & \textbf{BIB Rejection}  \\
\hline
$-90<t_{\mathrm{corr}}<51$~ps  & 95\%   & 92.5\% \\
$-90<t_{\mathrm{corr}}<64$~ps  & 98\%   & 91.8\% \\
$-90<t_{\mathrm{corr}}<73$~ps  & 99\%   & 91.3\% \\
\hline
\hline
\end{tabular}
\caption{Fraction of signal and BIB pixel clusters surviving various timing requirements }
\label{table:timing}
\end{table}

\subsection{Cluster Size Selection}
\label{sec:cut-based}
In addition to timing-based selections, the spatial and geometric properties of pixel clusters have also been investigated to further reduce pixel detector occupancy~\cite{angiraCHECKCITATION}. Following the \texttt{nominal} timing selection, a cluster-size requirement was optimized as a function of the detector position in $z$ to maintain a fixed signal efficiency of $95\%$. This additional selection reduces tracker occupancy from approximately 2000 to 850 hits/BX/$\mathrm{cm}^2$ for the innermost layer of the pixel detector. This selection results in a 57\% reduction in the overall data rate because large pixel clusters are preferentially removed~\cite{angiraCHECKCITATION}. This result demonstrates that geometric information can complement timing-based selections to achieve additional background rejection. 

\subsection{ML Algorithms}

We investigate the potential of machine learning to improve BIB rejection beyond fixed selections on pixel cluster size. We develop several neural networks that use pixel cluster shape information to distinguish BIB from signal clusters. The networks are quantized to enable implementation in digital electronics, providing a path toward on-chip inference and data reduction in a future front-end readout application-specific integrated circuit (ASIC).

A randomized hyperparameter scan is performed for each considered model architecture and each quantization configuration. Pareto-optimal models for each are selected based on the background rejection and an estimate of hardware resource usage. High-level synthesis (HLS) is used to estimate resource utilization for implementations targeting field-programmable gate arrays (FPGAs) and ASICs.


\subsubsection{Neural Network Architectures}

\begin{figure}[htbp]
    \centering
    \includegraphics[width=\linewidth]{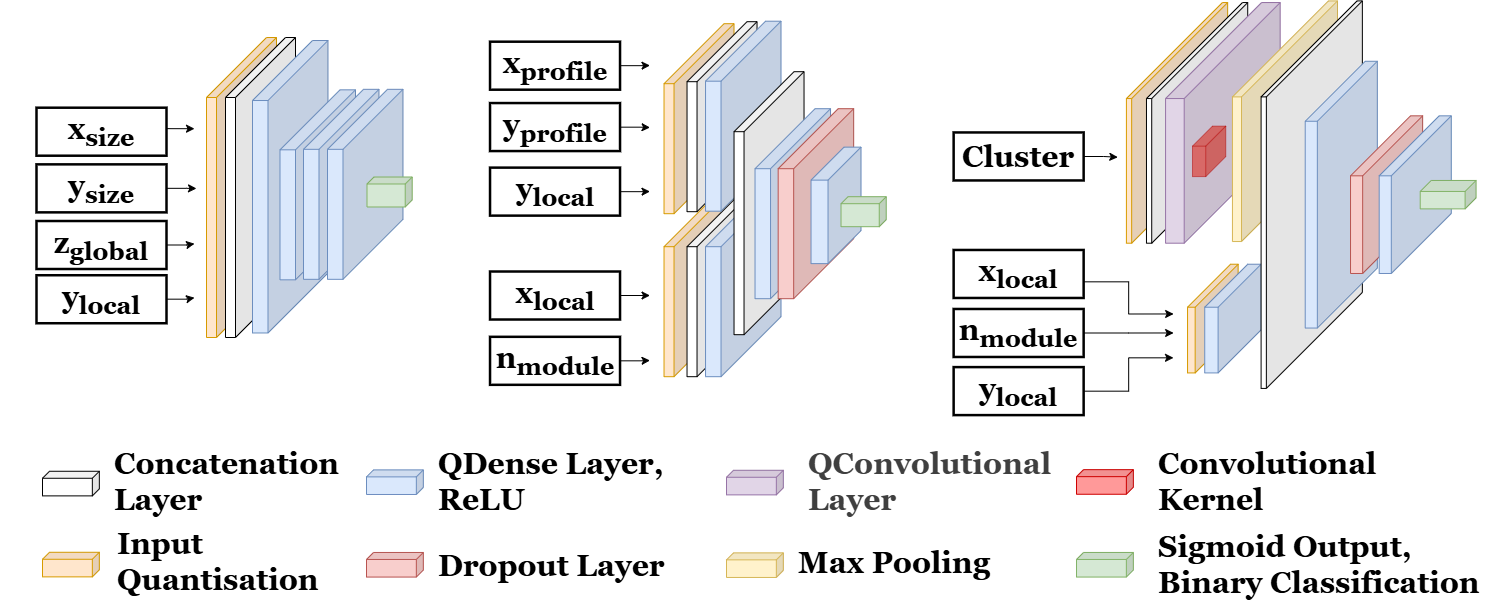}


    \caption{Schematic overview of the three classes of neural networks considered in this study: Model 1 uses compact engineered features (left), Model 2 uses one-dimensional charge profiles (middle), and Model 3 uses the full two-dimensional charge cluster (right). }
    \label{fig:models_side_by_side}
\end{figure}

Three main classes of models with varying levels of complexity are considered, as shown in Figure~\ref{fig:models_side_by_side}. Models use inputs described in Section~\ref{sec:datasetFeatures}, with  architectures chosen to investigate the tradeoffs between model complexity, hardware usage, and performance. In order to design filtering algorithms that are roughly orthogonal to timing-based selections, we do not train on measured or corrected hit time. 

\textbf{Model 1 (Minimal feature model):}  
This model uses four numerical inputs: \xsize, \ysize, \zglobal, and \ylocal. It is designed using only compact, engineered features to minimize the number of parameters. Separating \zglobal into \nmodule and \xlocal did not significantly improve performance due to the compact nature of the model. Inputs were concatenated and passed through a fully-connected dense network. 
    
\textbf{Model 2 (Profile-based model):}  
This model uses higher-dimensional engineered inputs while avoiding the full $21 \times 13$ \textit{charge cluster}. The inputs are \xprofile, \yprofile, \nmodule, \xlocal, and \ylocal. As in Model~1, the baseline architecture is a fully-connected (dense) network, with the profile vectors concatenated with the position coordinates as input.
    
\textbf{Model 3 (Full-cluster model):}  
This model uses the full $21 \times 13$ \textit{charge cluster}, \ylocal, \nmodule and \xlocal coordinates as inputs. A Convolutional Neural Network (CNN) architecture is used to process the cluster efficiently. The output of the CNN is concatenated with the position coordinates and passed through fully-connected dense layers. 



\subsubsection{Quantization and Training}

Models are implemented using TensorFlow \cite{tensorflow2015-whitepaper} with QKeras ~\cite{qkeras} to perform quantization-aware training. We considered 3-, 4-, 6-, 8-, and 10-bit quantization of the weights and biases for each architecture. Activations were quantized to 8-bits in all configurations. Inputs were quantized using two additional bits relative to the weight precision.

The batch size was fixed at 16,384, chosen as a compromise between training efficiency and model performance. Smaller batch sizes increased training time without substantial improvements to performance.
The learning rate was optimized as part of the hyperparameter search, and the Adam optimizer \cite{Kingma2014AdamAM} was used for training, along with the \texttt{binary crossentropy} loss function. We used the Keras \cite{chollet2015keras} \texttt{EarlyStopping} callback with a patience of 10 epochs to limit overtraining and reduce training time, although this criterion was rarely reached. Training was limited to a maximum of 150 epochs based on examination of training and validation loss curves. 


\subsubsection{Hyperparameter Search}
We used the built-in Keras tuner to perform a randomized hyperparameter scan for each model architecture and quantization, performing a training on the full dataset at each point. For each search, we identified configurations that minimize the number of model parameters and maximize \bibrej{99}, using the Pareto optimization algorithm~\cite{paretoIThinkThisIsOldestPaper?}. This optimization criteria allows us to explore the tradeoff between model performance and complexity, which is measured by the number of parameters and is expected to increase the resource usage in hardware implementations of the model. In total, we performed 15 Pareto scans, corresponding to three classes of models and five quantization levels each. The full set of hyperparameters explored is provided in Appendix~\ref{app:mlresults}, Table~\ref{tab:hyperparameter-search}.

\begin{figure}[htbp]
    \centering

    \begin{subfigure}[t]{0.49\linewidth}
        \centering
        \includegraphics[width=\linewidth]{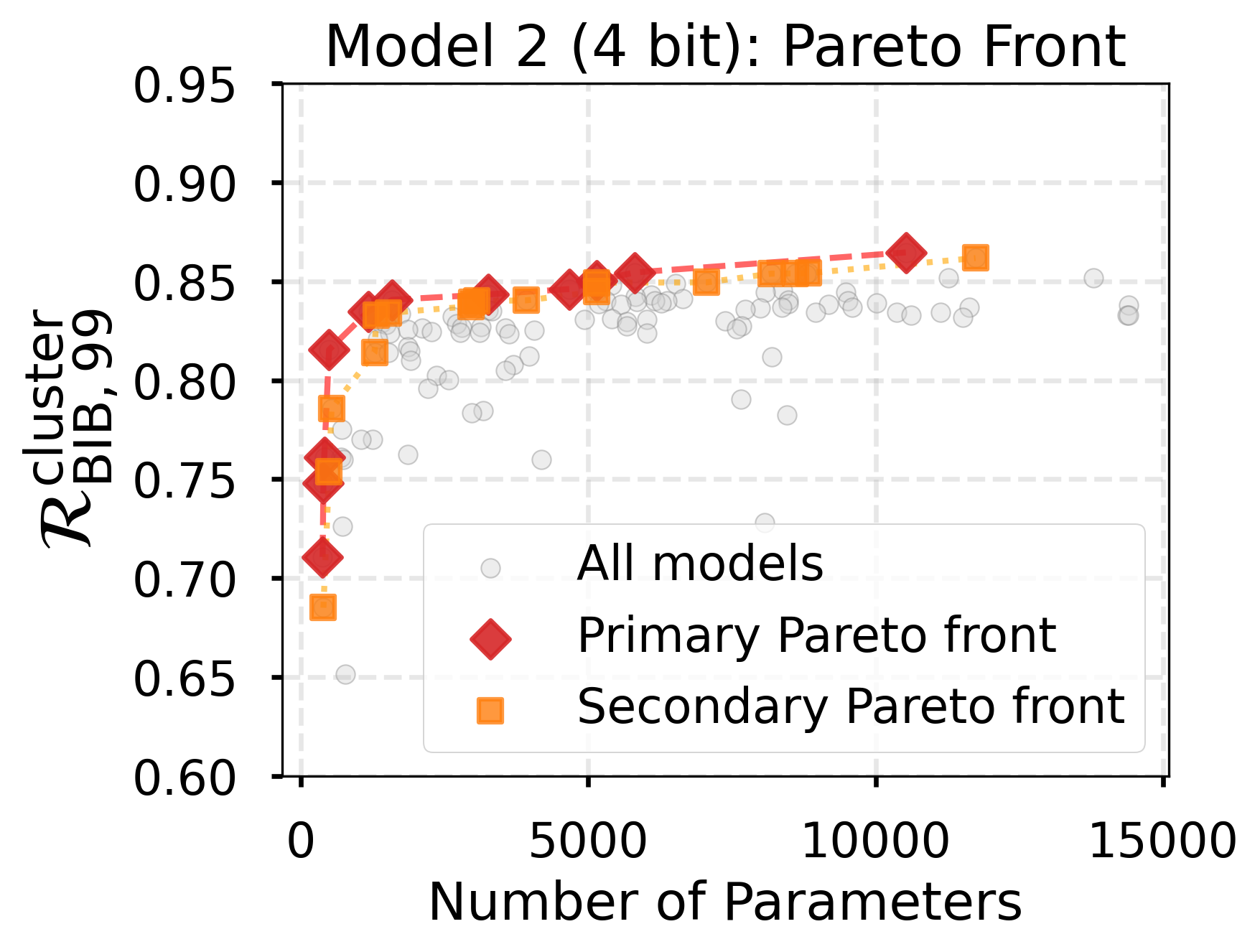}
        \caption{Model 2, 4 bit weights}
        \label{fig:}
    \end{subfigure}\hfill
    \begin{subfigure}[t]{0.49\linewidth}
        \centering
        \includegraphics[width=\linewidth]{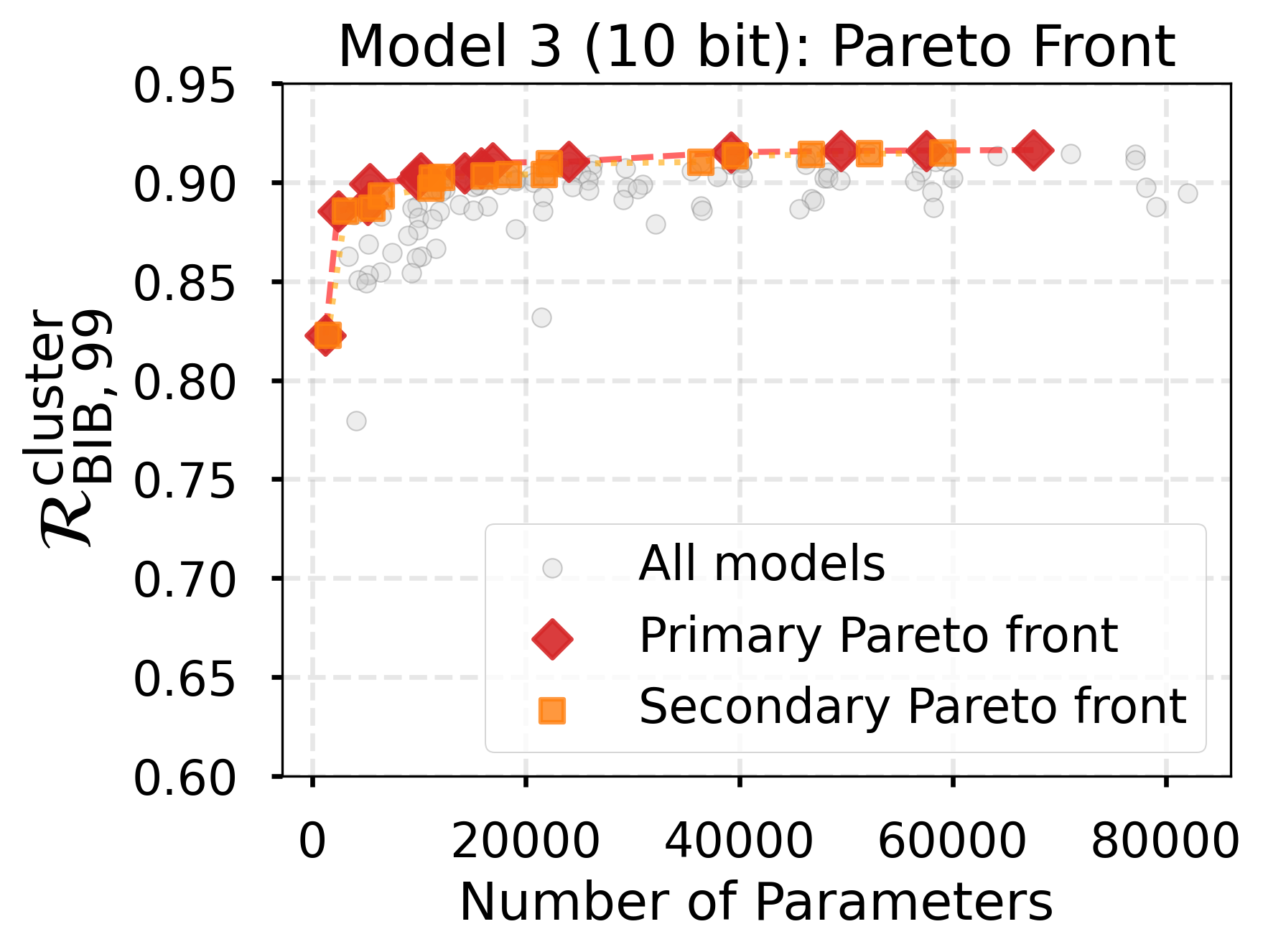}
        \caption{Model 3, 10 bit weights}
        \label{fig:}
    \end{subfigure}

    \caption{Example of 2 of the 15 Pareto fronts generated for each model at each quantization level, based on the background rejection at 99\% signal efficiency, \bibrej{99}, versus the number of parameters. Models from these Pareto fronts were selected for further study.}
    \label{fig:sampleHyperparamInitial}
\end{figure}


For Model 1, the hyperparameter search varied the number of layers (2–5) and neurons per layer (2–11). The learning rate was fixed to $10
^{-3}$ after tests on selected architectures showed no performance improvement from further variation. Ranges for layers and neurons were chosen based on preliminary studies that showed no improvement beyond these values. 
 


For Model 2, the hyperparameter search varied the learning rate and network architecture. The learning rate was scanned from $10^{-2}$ to $10^{-4}$. The architecture was defined by four parameters controlling the dense layer sizes: the numbers of neurons in the layers following the \nmodule + \xlocal and \xprofile + \yprofile + \ylocal concatenations (2–12 and 8–128 neurons, respectively), and the relative sizes of the two subsequent dense layers. The relative size was parameterized as a ratio to the preceding layer size and varied between 0.2 and 0.7. The dropout rate was evaluated as an additional parameter but was fixed to 0.08 for the final models.


For Model 3, the hyperparameter search varied the learning rate and network architecture. The learning rate was scanned from $10^{-2}$ and
$10^{-4}$. Architecture parameters included the convolutional layer size (2–10 neurons), kernel size, scalar input branch size (8–32 neurons), first dense layer size (8–128 neurons), and the ratio of the second to first dense layer sizes (0.2–0.8). The kernel size was ultimately fixed to $3\times3$, and the dropout rate to 0.08 after initial exploration. The scalar input branch was merged with the convolutional output before the final two dense layers.


A randomized hyperparameter scan was performed for each model and each quantization level. Two representative hyperparameter scans are shown in Figure~\ref{fig:sampleHyperparamInitial}. For each scan, models along the Pareto front, which had optimal performance 
with minimal parameter count, were selected for further study of hardware resource usage. For Models 1 and 2, we also consider the secondary Pareto fronts, constructed by removing models lying on the primary Pareto front and repeating the Pareto optimization algorithm. In total, the 25 Pareto fronts yielded 329 models for further study. 

\begin{figure}[htbp]
    \centering
    \includegraphics[width=0.48\linewidth]{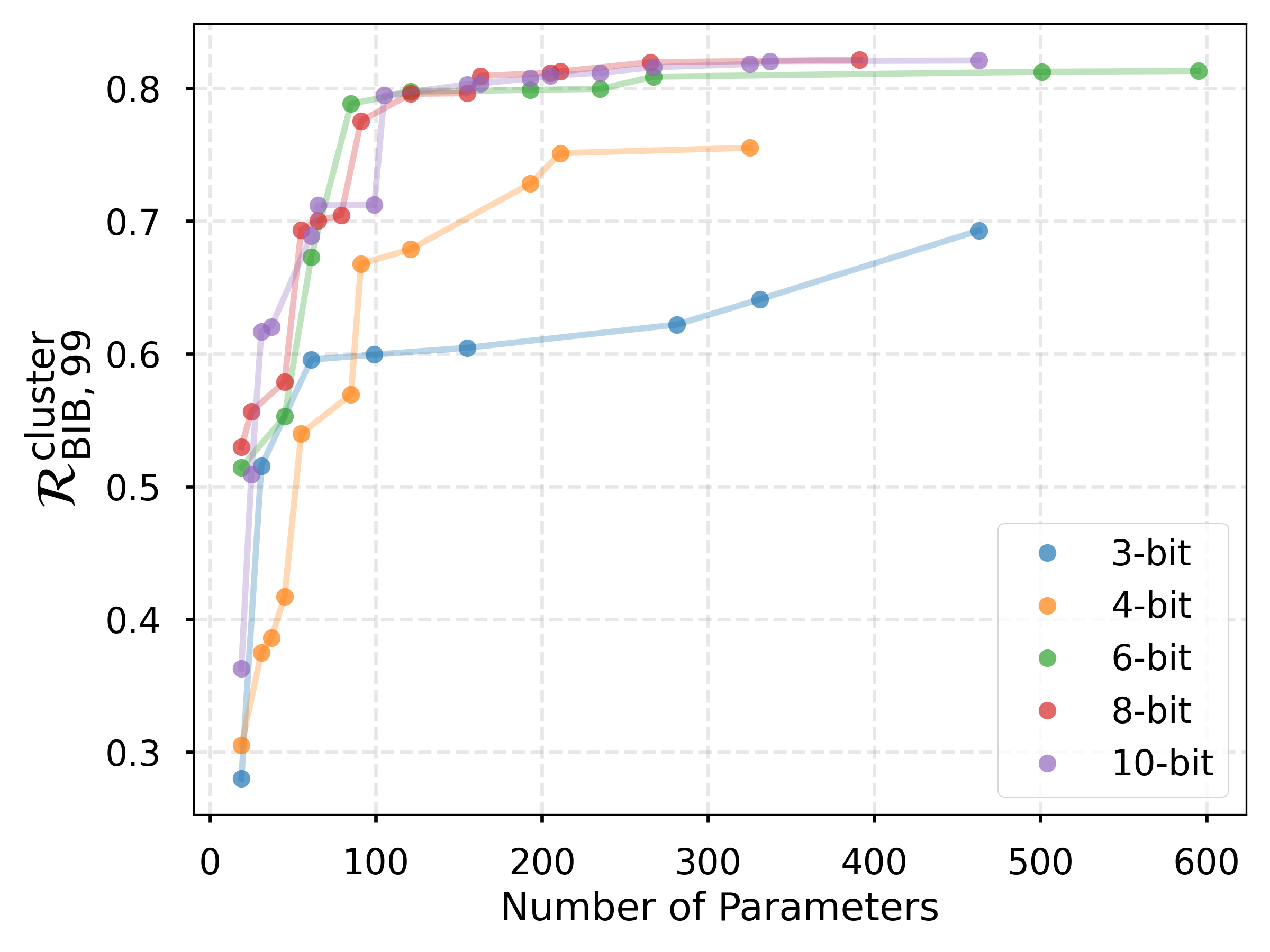}
    \includegraphics[width=0.48\linewidth]{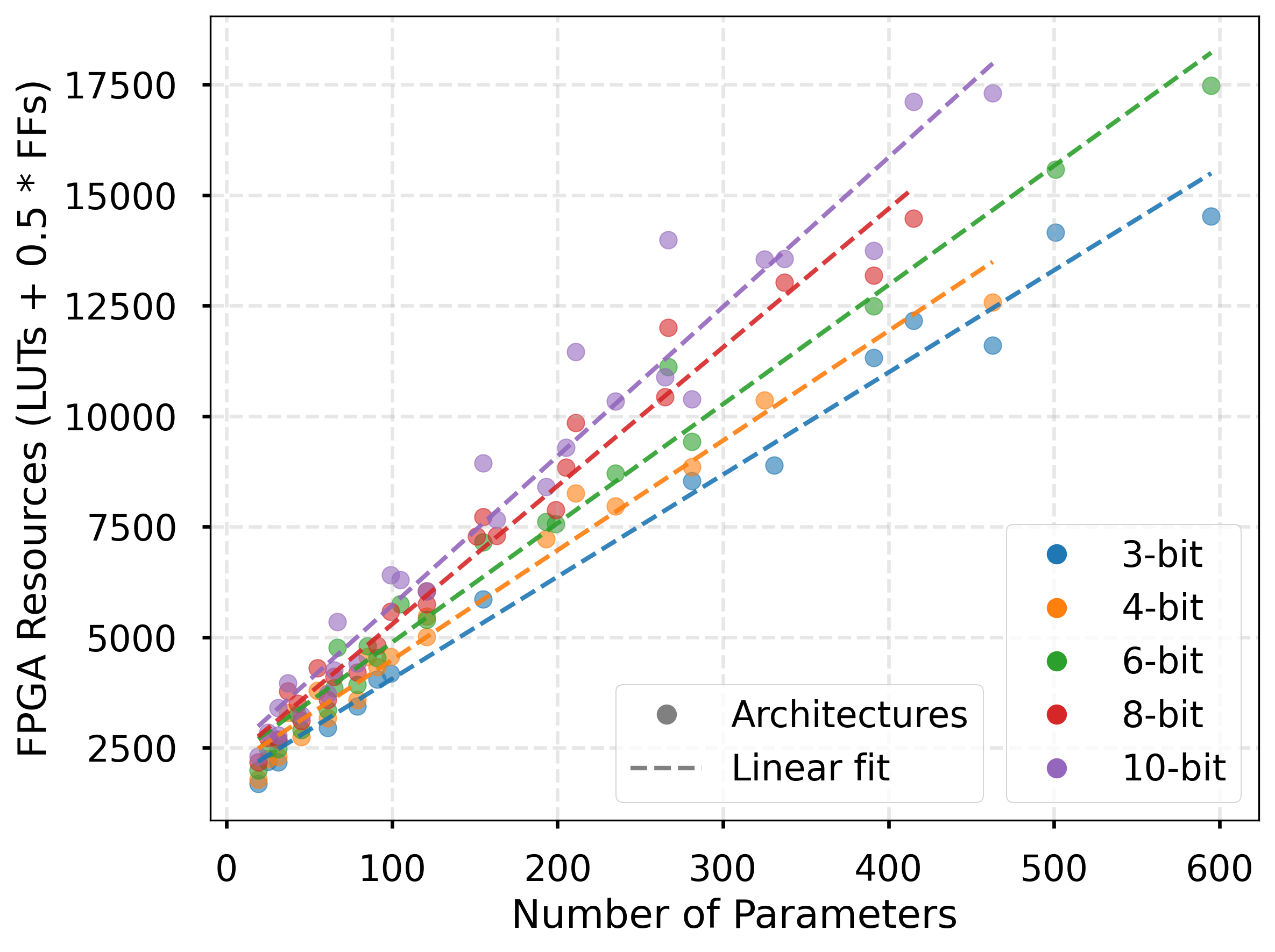}
    \caption{(left) Primary Pareto fronts for the Model 1 hyperparameter scan at each quantization. (right) Estimated FPGA resource usage for Model 1 architectures obtained from the Vitis synthesis (v-synth) stage of \texttt{hls4ml}, as a function of the number of parameters. 
    }
    \label{fig:pareto_bits_hwresources}
\end{figure}
\subsubsection{Hardware Resource Utilization and Model Selection}

We used high level synthesis (HLS) to estimate resource usage for these 329 models. \texttt{QKeras} models were first converted with \texttt{hls4ml} \cite{hls4ml_overview_latest,fastml_hls4ml,hls4ml_cnn_1,hls4ml_cnn_2} into C++ implementations. The \texttt{Vitis HLS} backend \cite{vitis} was then used to synthesize an FPGA implementation. Finally, we explored the \texttt{Siemens Catapult  HLS} backend \cite{catapult} to estimate approximate hardware usage in a preliminary ASIC implementation. 

We first used the C-synthesis stage of \texttt{hls4ml} (c-synth) to estimate FPGA resource utilization for each of the 329 models discussed above. While C-synthesis resource estimates are less accurate than those from an RTL implementation, they enable rapid resource estimation across all 329 models. Synthesis was constrained to disable digital signal processors (DSPs). We characterized the approximate hardware cost as the sum of the number of lookup tables (LUTs) and half the number of flip-flops (FFs), denoted as ``LUTs + 0.5 $\times$ FFs". The coefficient of 0.5 was chosen based on logical units of the target FGPA considered~\cite{amd7datasheet}. Modifying this coefficient between 0 and 2 did not impact the final model selection, so 0.5 was chosen for consistency.

We then constructed a cross-model Pareto front to identify model configurations that provide the optimal trade-off between performance and approximate hardware resource usage from c-synth for further study. These Pareto-optimal configurations were then synthesized with \texttt{hls4ml} using the \texttt{Vitis HLS} backend to obtain resource estimates from a realistic RTL implementation (v-synth). 

The relationship between model size, quantization, and estimated hardware resource usage is illustrated in Figure~\ref{fig:pareto_bits_hwresources} for for Model 1. Primary Pareto fronts are shown for different quantization levels, and total hardware resource usage is shown as a function of the number of parameters. For smaller models, including all instances of Model 1, resource usage increases approximately linearly with model size, with steeper slopes at higher quantization precision due to the additional resources required for higher-precision weights and biases. For Model 2, this relationship appears to break down when the number of trainable parameters exceeds about 6,000. 

Finally, ASIC resource estimates were obtained for a few representative model configurations from a preliminary synthesis using \texttt{Catapult HLS} with the publicly available \texttt{SAED32 EDK} library~\cite{edkCHECKIFRIGHTCITATION}, providing first-pass estimates for a hypothetical 32 nm CMOS implementation. These estimates serve as a relative comparison metric and are not representative of a manufacturable ASIC implementation.

\begin{figure}[h]
    \centering
    \includegraphics[width=0.98\linewidth]{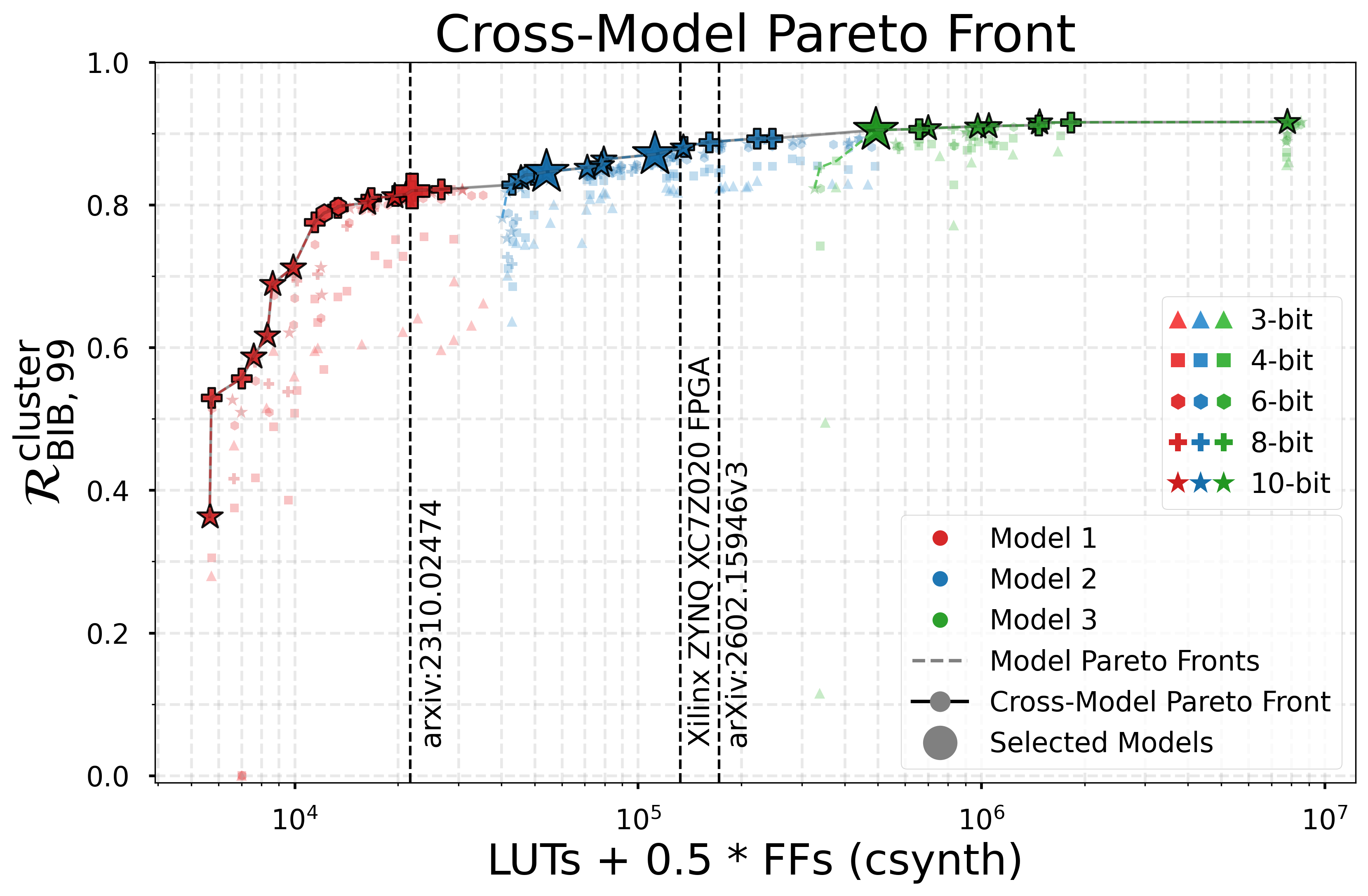}
    \caption{Background rejection at 99\% signal efficiency, \bibrej{99}, versus the estimated hardware usage on an FPGA, LUTs + 0.5 * FFs, for the 329 model configurations identified from hyperparameter scans. 
    A Pareto front is selected across all models, and 4 configurations are selected for further study. Resource estimations of \smartpixels neural networks~\cite{filtering,regression} and the FPGA featured on a Xilinx Pynq Z2 
    are shown for reference.}
    \label{fig:crosspareto}
\end{figure}
\section{Results}

The cross-model Pareto front with C-synthesis resource estimations is shown in Figure~\ref{fig:crosspareto}. 
For reference, the resources of a Xilinx Pynq Z2 are included, as well as resource estimations for neural network architectures discussed in \cite{filtering} and the slim model in \cite{regression}.

Model performance generally increases with estimated resource usage. The performance of Models 1, 2 and 3 plateau at increasingly higher background rejection factors. However, the increased architecture complexity results in order of magnitude increases in hardware usage.  Quantization level alone does not dictate performance. Larger models with less precision can have comparable performance to smaller models with higher precision.

Four configurations were selected for further study. One Model 1 and two Model 2 architectures were chosen for their similarity in resource usage with the \smartpixels filtering and regression algorithms, which have been or are planned to be manufactured in 28 nm CMOS prototype ASICs~\cite{filtering,regression,testing}. The smaller Model 2 architecture was selected as an intermediary between the filtering and regression algorithms, and its detailed performance is presented below. The smallest Model 3 configuration on the Pareto front was also selected for a point of comparison.  

The estimated FPGA and ASIC resource usage for these four configurations are summarized in Table~\ref{table:bkrejAndHW}, with additional performance results provided in Appendix~\ref{app:mlresults}. 
We conclude that certain configurations of Models 1 and 2 are well suited for implementation on an ASIC, while Model 3 would require too many resources.

\begin{table}[h]
\centering
\begin{tabular}{l|l|c|cc|c}
\hline
\multicolumn{2}{l|}{\textbf{Model Architecture}} & \textbf{Model 1} & \multicolumn{2}{c|}{\textbf{Model 2}} & \textbf{Model 3} \\
\hline
\multicolumn{2}{l|}{\textbf{Weight Quantization}} & 8-bit & 10-bit & 10-bit & 10-bit \\
\hline
\multicolumn{2}{l|}{\textbf{Number of Parameters}} & 265 & 711 & 2,637 & 14,241 \\
\hline
\multirow{3}{*}{\begin{tabular}{@{}l@{}}\textbf{Cluster}\\\textbf{Rejection} \\\textbf{Rate}\end{tabular}} 
& \bibrej{95}   & 86.9 \% & 88.9 \% & 91.8 \% & 93.8 \% \\ 
& \bibrej{98}  & 84.7 \% & 86.8 \% & 89.5 \% & 92.1 \% \\ 
& \bibrej{99}  & 82.0 \% & 84.6 \% & 87.1 \% & 90.5 \% \\
\hline
\multirow{2}{*}{\begin{tabular}{@{}l@{}}\textbf{Data}\\\textbf{Reduction} \end{tabular}} 
 & \bibred{99}  & 88.4 \% & 90.4 \% & 91.8 \% & 94.4 \% \\
 & \bibred{99} + $[3\sigma_t, 5\sigma_t]$  & 98.5 \% & 98.7 \% & 98.9 \% & 99.3 \% \\
\hline

\hline
\multirow{6}{*}{\begin{tabular}{@{}l@{}}\textbf{FPGA}\\\textbf{Resource} \\\textbf{Usage}\\\textbf{Programmable}\\\textbf{Weights}\\ \textbf{(v-synth)}\end{tabular}} 
    & LUT      & 5,588   & 11,465  & 20,026 & 145,551  \\
    & FF       & 5,413   & 12,337  & 23,381  & 104,176  \\
    & DSP      & 0       & 0       & 0       & 0      \\
    & BRAM     & 0       & 0       & 0       & 0 \\
    & Clock Period [ns] &10.9 &11.4 &11.5 & 11.9 \\
    & Target Clock [ns] &12 & 12   & 12      & 12  \\ 
    & Latency** [$\mu$s]   & 1.7*  &  4.3*  & 15.6*    & 164* \\
    & Initiation Interval** & 32 & 280  & 700       & 6553 \\
\hline
\multirow{2}{*}{\begin{tabular}{@{}l@{}}\textbf{ASIC}\\\textbf{Resources}\\ \textbf{Estimates}\end{tabular}} 
    & Area Score [$\unit{\milli\meter}^2$] & 0.140  & 0.586  & 1.900  & 8.394 \\ 
    
    & Clock Period [ns] &10  & 10   & 10    & 10  \\
    & Latency [ns]   & 200  & 250  & 260   & 19,460  \\
\hline
\end{tabular}
\caption{Comparison including quantization, background rejection, FPGA v-synth resources, and ASIC metrics of four representative configurations.
*FPGA implementations with programmable weights do not have a deterministic pipeline, so the maximum latency is presented. **This implementation includes loading weights and biases with each inference, which unrealistically increases the latency and initiation interval.}
\label{table:bkrejAndHW}
\end{table}


The implementations presented in Table \ref{table:bkrejAndHW} all have programmable weights and biases. On an ASIC, programmable weights (and biases) allow reconfigurability after the ASIC has been manufactured. \texttt{Catapult} maps programmable weights to registers, so the latencies presented are realistic for an inference of a single input. For consistency, the FPGA implementation is also presented with programmable weights and biases. However, the HLS implementation for FPGA loads all the weights with every inference, which unrealistically increases the latency. A practical FPGA implementation would have non-programmable weights because the entire design could be reprogrammed, which would decrease the latency on an FPGA to be comparable to the ASIC latencies. Further details on alternate FPGA implementations are available in Appendix \ref{app:mlresults}.

\begin{figure}
    \centering
    \includegraphics[width=0.45\linewidth]{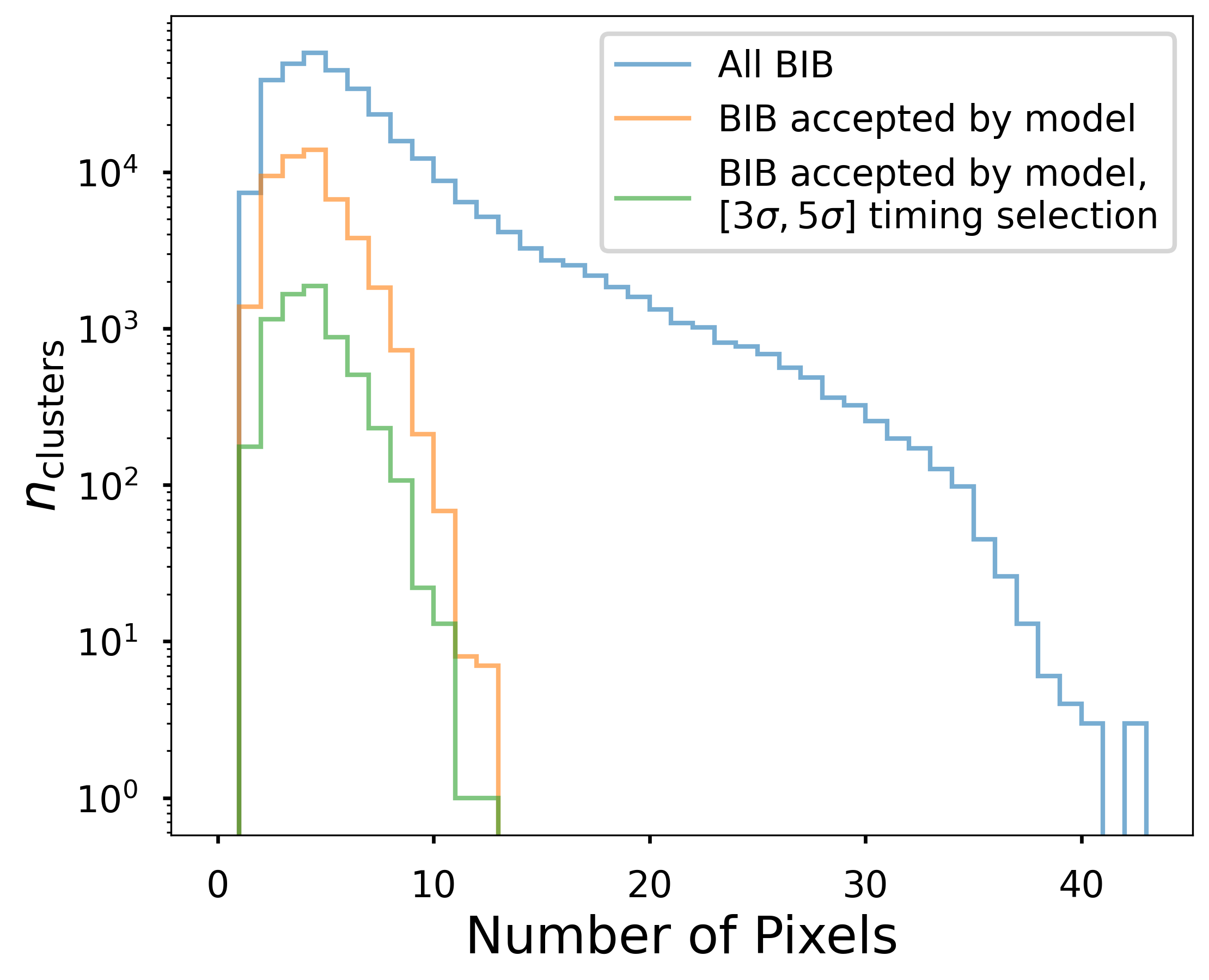}
    \includegraphics[width=0.45\linewidth]{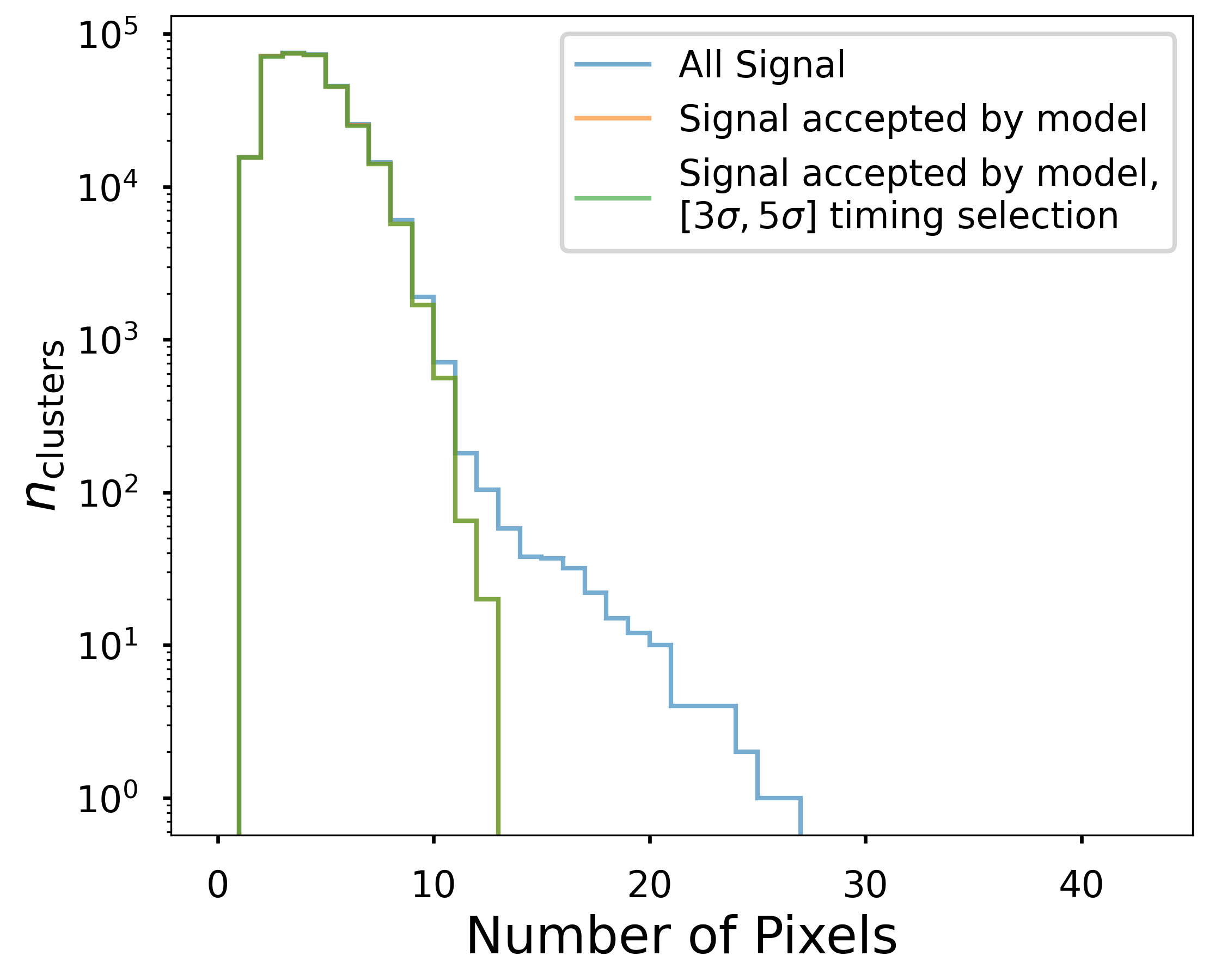}
    \caption{The number of pixels per cluster for BIB (left) and Signal (right)  accepted and rejected by the selected Model 2 architecture.}
    \label{fig:model57Npix}
\end{figure}

The ultimate goal of these networks is to reduce the BIB data rate in order to enable pixel front-end readout. In addition to evaluating the fraction of BIB clusters rejected, we also evaluate the corresponding reduction in data rate, accounting for the number of pixel hits in each cluster. Figure~\ref{fig:model57Npix} shows, for a representative Model 2 architecture, that BIB filters preferentially reject larger clusters, translating the \bibrej{99} to an even larger reduction in the data rate, \bibred{99}. Since BIB is expected to dominate tracker occupancy at a 10 TeV Muon Collider, this provides a first-order estimate of the total data reduction. All four considered models achieve approximately an order of magnitude data reduction.

\begin{figure}[htbp]
    \centering

    \begin{subfigure}[t]{0.97\linewidth}
        \centering
\includegraphics[width=\linewidth]{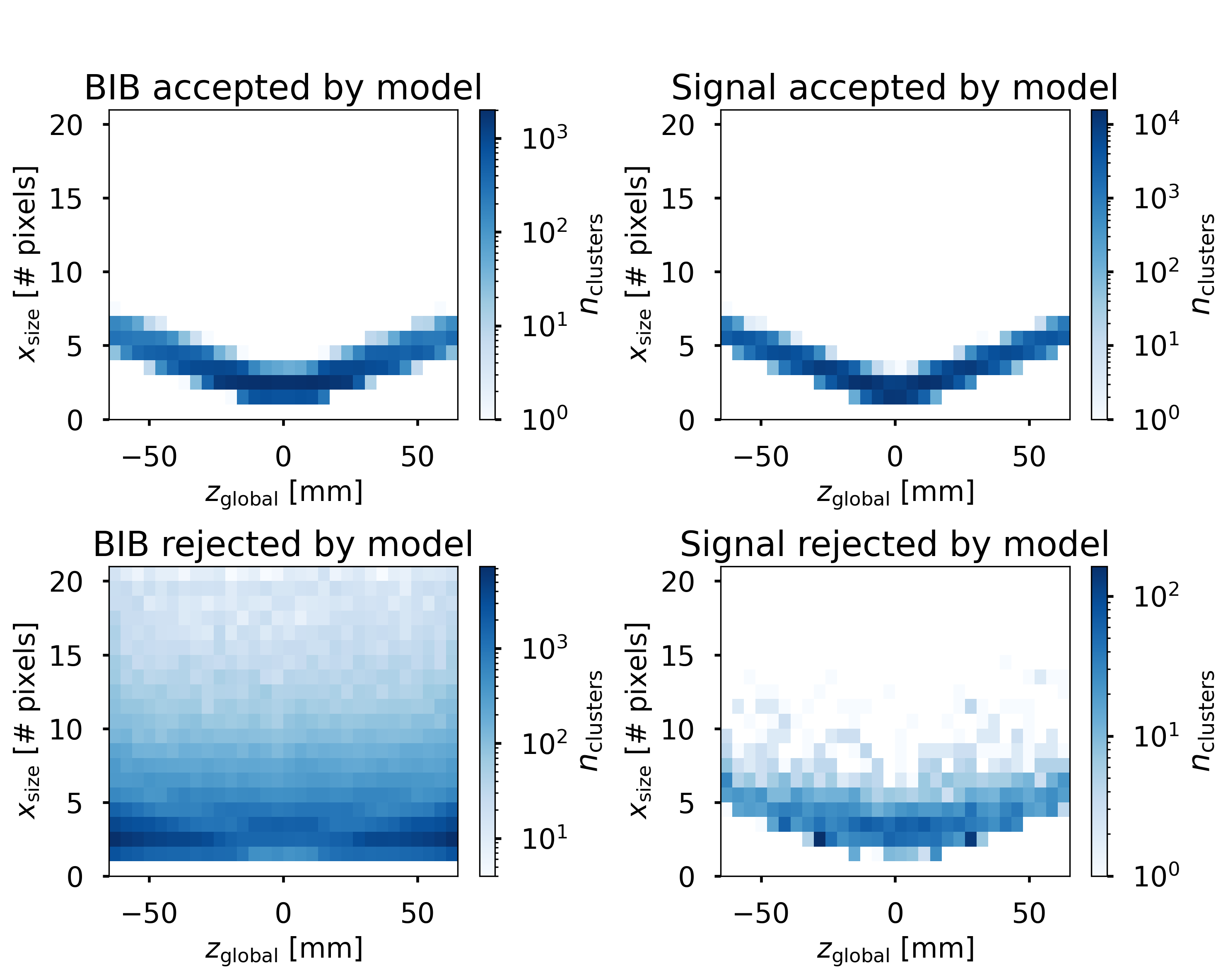}
        \label{fig:}
    \end{subfigure}\hfill
    \caption{Distributions of the \xsize versus \zglobal for BIB (left) and Signal (right) clusters accepted (top) and rejected (bottom) by the selected Model 2 architecture.}
    \label{fig:model57Performance}
\end{figure}

\begin{figure}
    \begin{subfigure}[t]{0.49\linewidth}
        \centering
        \includegraphics[width=\linewidth]{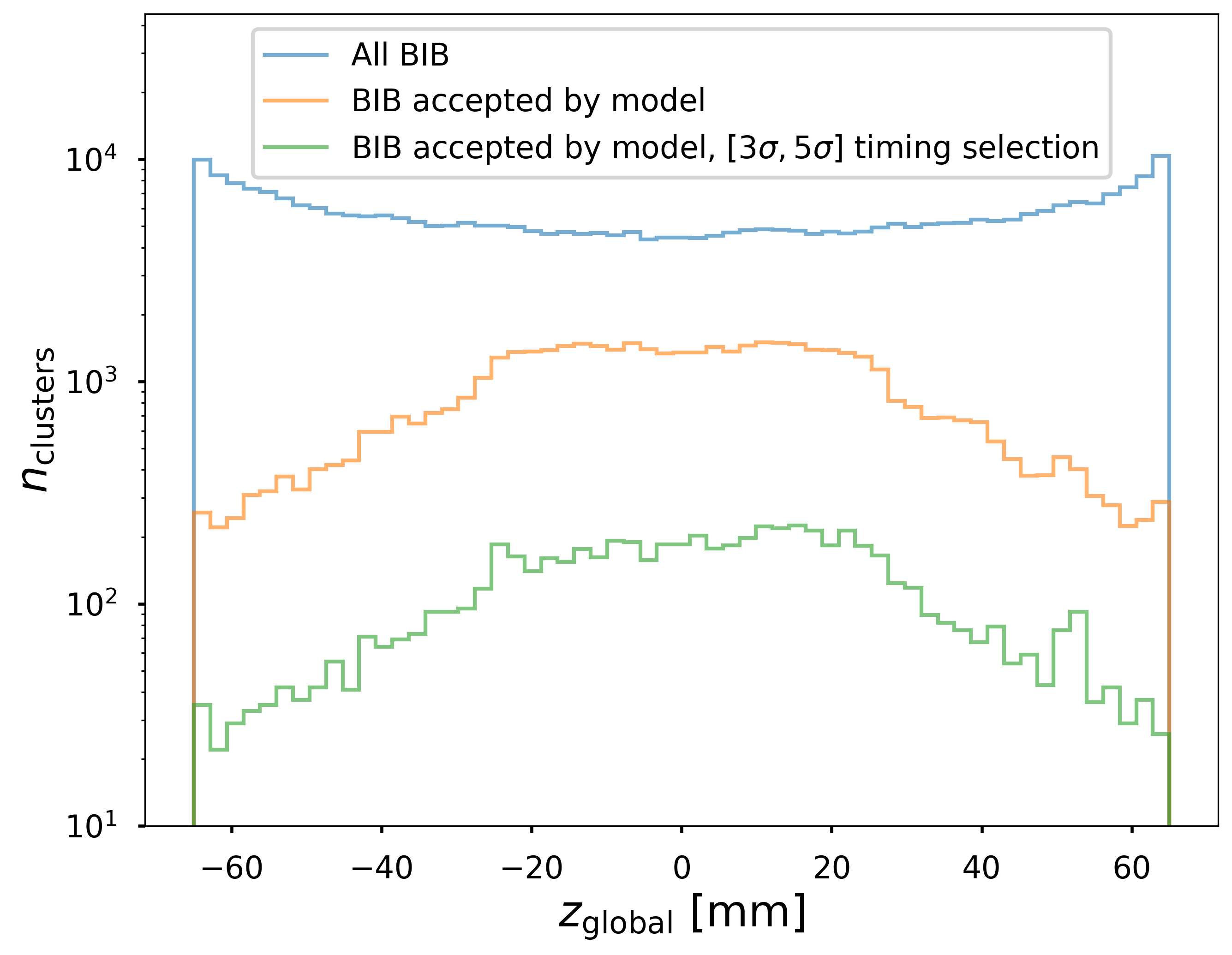}
        \label{fig:}
    \end{subfigure}\hfill
    \begin{subfigure}[t]{0.49\linewidth}
        \centering
    \includegraphics[width=\linewidth]{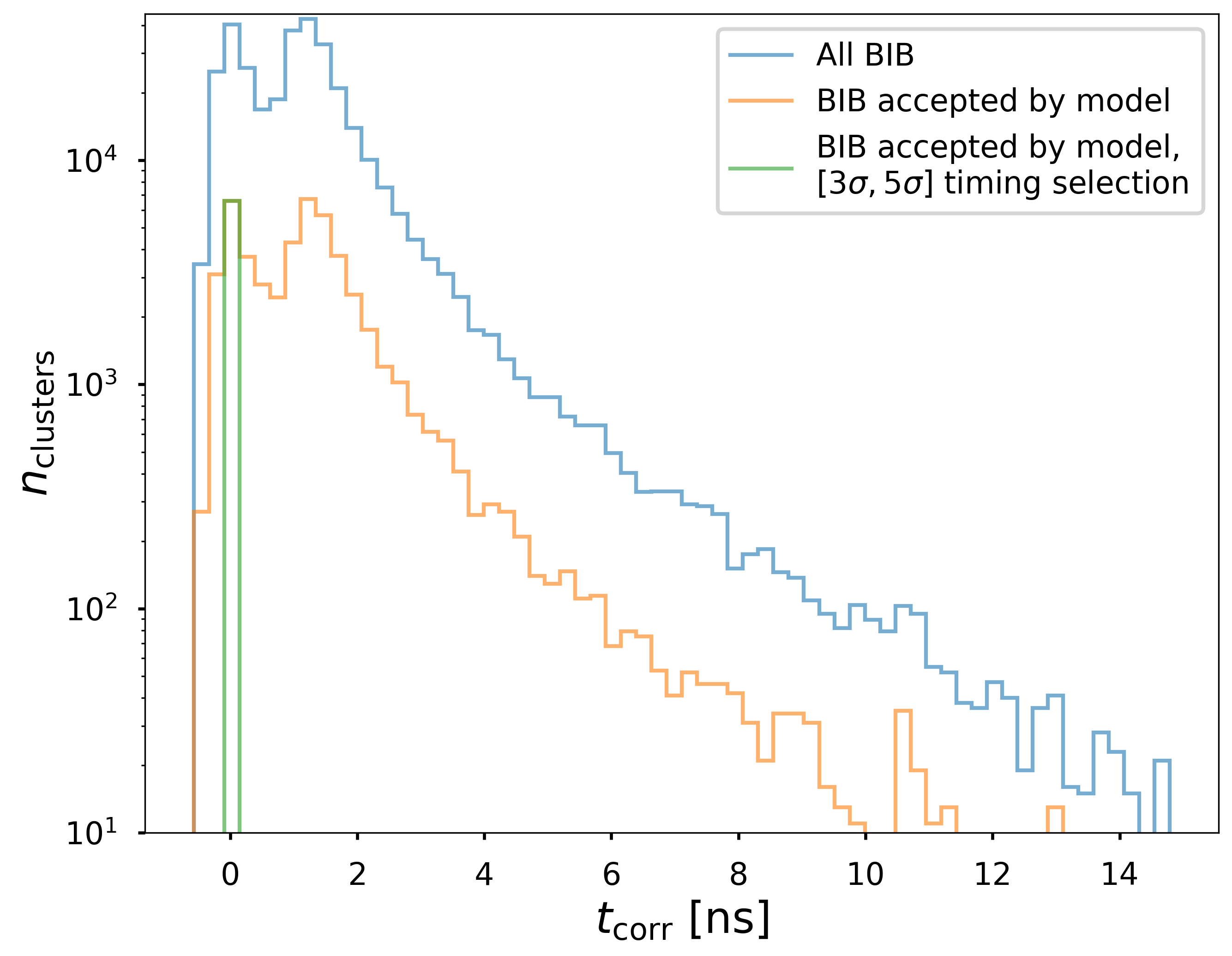}
        \label{fig:}
    \end{subfigure}
    \caption{Histogram of \texttt{z-global} (left) and the adjusted arrival time of clusters (right) for BIB clusters accepted and rejected by the selected Model 2 architecture. The impact of the $[3\sigma, 5\sigma]$ timing window is also shown.}
    \label{fig:model57time}
\end{figure}

Figures \ref{fig:model57Performance} and \ref{fig:model57time} show the properties of clusters that are accepted and rejected by the classifier for the selected Model 2. The classifier preferentially selects BIB clusters with an \xsize consistent with signal clusters for a given \zglobal, as shown in Figure~\ref{fig:model57Performance}. Furthermore, the classifier preferentially rejects BIB clusters at large \textbar{\zglobal}\textbar, where the incident angles of collision and BIB particles differ most strongly, as shown in Figure \ref{fig:model57time}. Figure~\ref{fig:model57time} also shows the corrected hit time for accepted and rejected BIB  clusters. The timing distributions show that the classifier rejects BIB across the full timing spectrum, demonstrating that cluster shape provides information largely complementary to hit timing.

The classification dependence on \zglobal indicates that the achievable background rejection depends on detector location. In practice, this dependence could be reduced or eliminated by varying the cut used in each module or by retraining the model for different detector regions and deploying common hardware with different network weights and biases to each module. The second approach would remove the need to encode large variations in the local angular distributions across the detector in the model weights.





\section{Conclusions and Future Work}

We have demonstrated that quantized neural networks using pixel-cluster shape and position as inputs can be used to filter BIB in the innermost pixel layer of a 10 TeV Muon Collider detector. Three candidate architectures achieve 88.4–94.4\% data rate reduction at 99\% signal efficiency. This result substantially exceeds data rate reductions achieved with fixed selections on cluster-size. This demonstrates that on-chip neural inference is well-suited to enable on-detector data reduction at a Muon Collider, supporting further development of the \smartpixels concept.

We find that both model performance and hardware cost increase with architecture complexity. The estimated ASIC area increases by two orders of magnitude between Model 1 and 3, and only translates to a $10\%$ improvement in data rate reduction. Models 1 and 2 are the most promising candidates for implementation on a front-end readout ASIC, achieving data rate reductions of 88–90\% at 99\% signal efficiency with ASIC area estimates below $1~\si{\milli\meter}^2$ and latencies of 160–250 ns. Model 3 reaches the highest background rejection but its estimated resource usage is prohibitive for a real-time front-end ASIC application.

Most importantly, we demonstrate that neural networks based on pixel cluster shapes provide an orthogonal handle to precision timing information. The neural network models considered here provide reductions in data rate comparable to those attained with precision timing. 
A $[3\sigma_t, 5\sigma_t]$ requirement on the hit time of arrival, assuming $30\ \si{\pico\second}$ resolution, reduces data rates by 85.6\% while Models 1 and 2 reduce data rates by 88-92\% for 99\% signal efficiency.
Combining model rejection with a $[3\sigma_t, 5\sigma_t]$ timing selection improves data reduction to 98.5-98.9\%.
A looser timing requirement and cluster-based neural network could plausibly be combined for on-detector data reduction, allowing for a less ambitious target timing resolution.

Several aspects of the dataset and detector model can be improved in future studies. The single muon dataset provides a useful approximation of particles of interest from collisions, but simulations of Standard Model and Beyond Standard Model signatures would enable a more realistic assessment. In a similar vein, an updated BIB sample with refinements to the accelerator lattice and machine–detector interface is available for study~\cite{maia}.

The BIB sample considered here does not include incoherent pair production (IPP). IPP results in extremely low-$p_{\mathrm{T}}$ particles, similar to muon decays from the beam, but these particles originate at the interaction point. The vast majority of hits due to IPP are expected to occur in the innermost layer of the vertex detector, and result in similarly long pixel clusters to our BIB sample. The similarity of these clusters suggests that IPP could also be rejected by the neural networks considered here, provided IPP is included in the training and evaluation samples.

Further simulation should account for additional detector effects such as sensor geometry, variations in electronic noise, and radiation damage. We assume that upstream front-end logic can provide a region of interest (ROI) containing a single isolated cluster. This logic would need to be updated for the high-occupancy environment of a Muon Collider, which will produce nearby and overlapping clusters. Effects associated with particle hits near module boundaries should also be investigated. The simulation could also be extended to multiple pixel layers, including the endcaps, to assess how performance evolves throughout the tracker.

The ASIC area scores presented in Table~\ref{table:bkrejAndHW} are preliminary estimates from the open source \texttt{SAED32 EDK} library. If an ASIC design is selected for tapeout, a commercial library would be chosen to generate a realistic implementation, with additional optimizations applied to meet power and latency constraints. 
A realistic ASIC would also target the final tracker design.
  
Alternative network architectures and technologies should also be explored to identify designs that maximize physics performance while satisfying realistic area, power, and latency constraints. Finally, the combined performance of shape-based filtering and precision timing should be quantified directly. 

\acknowledgments

We are grateful to the International Muon Collider Collaboration, whose software development was crucial to enable this work. We also acknowledge the Fast Machine Learning community whose multi-domain expertise was important for the development of this project. We are also grateful to Mary Heintz for her support in configuring and maintaining the computing resources used in this work.

We acknowledge the use of AI/LLM and code-completion tools to assist with prototyping and writing code. All code written by AI was extensively reviewed by human authors. We also acknowledge the use of AI/LLM tools to assist with formatting and typesetting of Latex. All content is written exclusively by human authors without AI assistance. 



KFD, BRR, BR, and DA are supported by the Simons Foundation through award SFI-MPS-T-MPS00010555. KFD and EH are supported by the NSF CAREER Program through award 2443370, and KFD is additionally supported by the Neubauer Family Assistant Professor Program. TNY, RM, and ML are supported by the University of Chicago’s Quad Undergraduate Research Scholar program and the Jeff Metcalf Internship program.

DB, GDG, FF, AG, LG, JH, RL, BP, and CS are supported by Fermi Forward Discovery Group, LLC under Contract No. 89243024CSC000002 with the U.S. Department of Energy, Office of Science, Office of High Energy Physics. GDG and LG are partially supported under DOE project ``HAAI: Designing Smart Detectors with a ML-to-Silicon Platform'' (LAB 24-3305). GDG and BP are also supported by the U.S. Department of Energy, Office of Science, Offices of High Energy Physics and Advanced Scientific Computing Research, as part of the Co-design and Heterogeneous Integration in Microelectronics for Extreme Environments (CHIME) Microelectronics Science Research Center (MSRC) (DOE Lab Announcement No. LAB 24-3320).

NT is also supported by the DOE Office of Science, Office of Advanced Scientific Computing Research under the “Real-time Data Reduction Codesign at the Extreme Edge for Science” Project (DE-FOA-0002501).

JD and BW are supported by the DOE Early Career Research Program award DE-SC0026236.

MS is supported by NSF-PHY award 2012584. 

DS is supported by the Visiting Scholars Award Program of the Universities Research Association. CM and DS are supported by NSF-PHY award 2513237. ART and CM are supported by DOE DE-SC0023715 (482662 / EB645).

MSN is supported through NSF cooperative agreement OAC-2117997 and the DOE Office of Science, Office of High Energy Physics, under Contract No. DE-SC0023365.

AB is supported by the Schmidt Sciences Foundation.


\bibliographystyle{JHEP}
\bibliography{biblio.bib}

\newpage
\appendix
\label{sec:appendix}

\section{Appendix A: Dataset Plots}
\label{app:dataset}
\begin{figure}[h]
    \centering
    \includegraphics[width=0.95\linewidth]{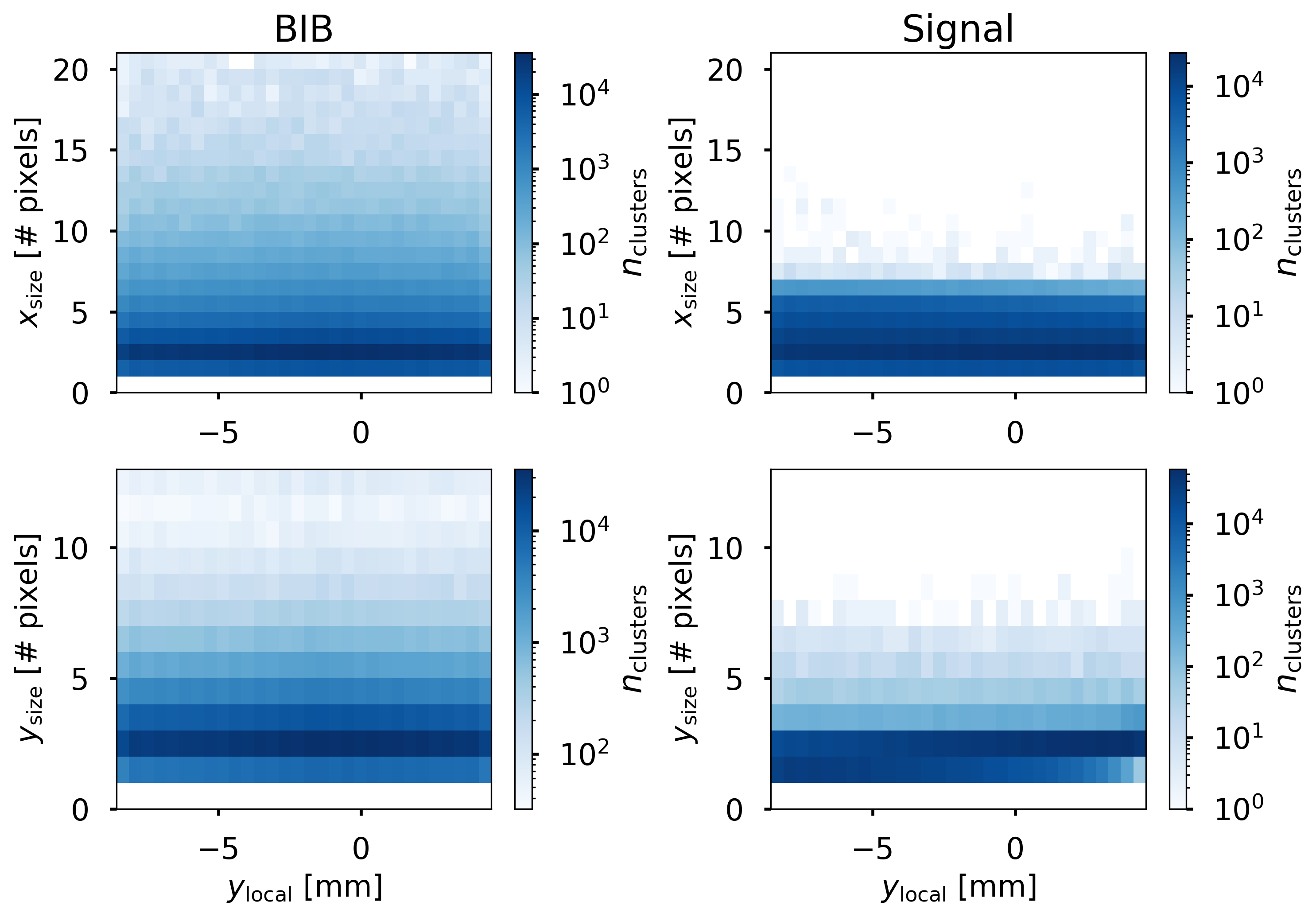}
    \caption{Distribution of the \xsize and \ysize versus \ylocal.}
    \label{fig:xysize_ylocal}
\end{figure}

\begin{figure}[h]
    \centering
    \includegraphics[width=0.95\linewidth]{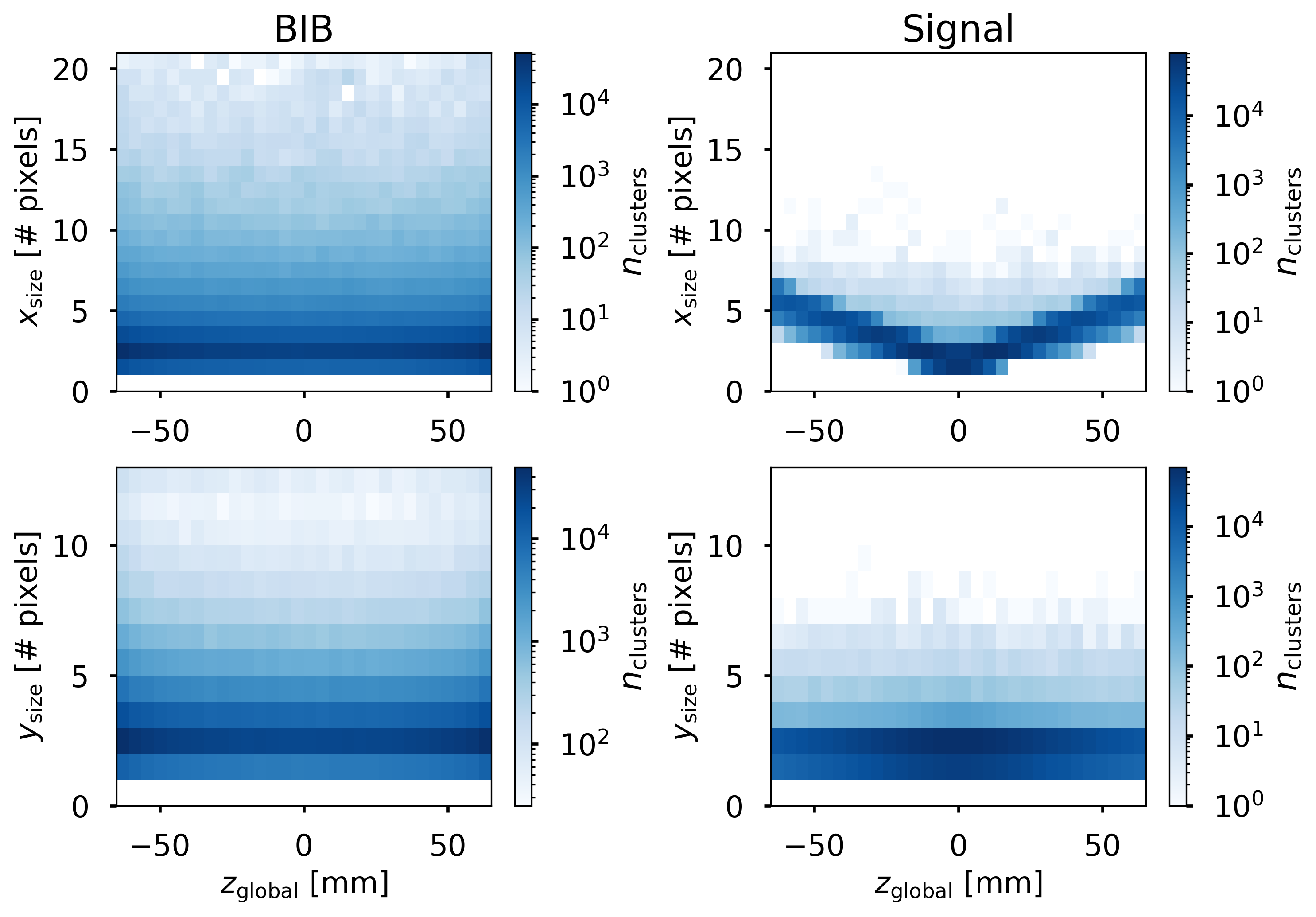}
    \caption{Distribution of the \xsize and \ysize versus \zglobal.}
    \label{fig:xysize_zglobal}
\end{figure}

Figures \ref{fig:xysize_ylocal} and \ref{fig:xysize_zglobal} show the distributions of \xsize and \ysize versus the longitudinal coordinates \zglobal and the perpendicular module coordinate \ylocal. As discussed above and shown in Figure \ref{fig:xsizezglobal}, the longitudinal size and coordinate are correlated for signal but not for BIB. Similarly, the \ysize is correlated with \ylocal for signal but not BIB, as particles arriving from the origin which curve more hit the farther end of the module and produce clusters with higher \ysize.  

\begin{figure}
    \centering
    \includegraphics[width=0.9\linewidth]{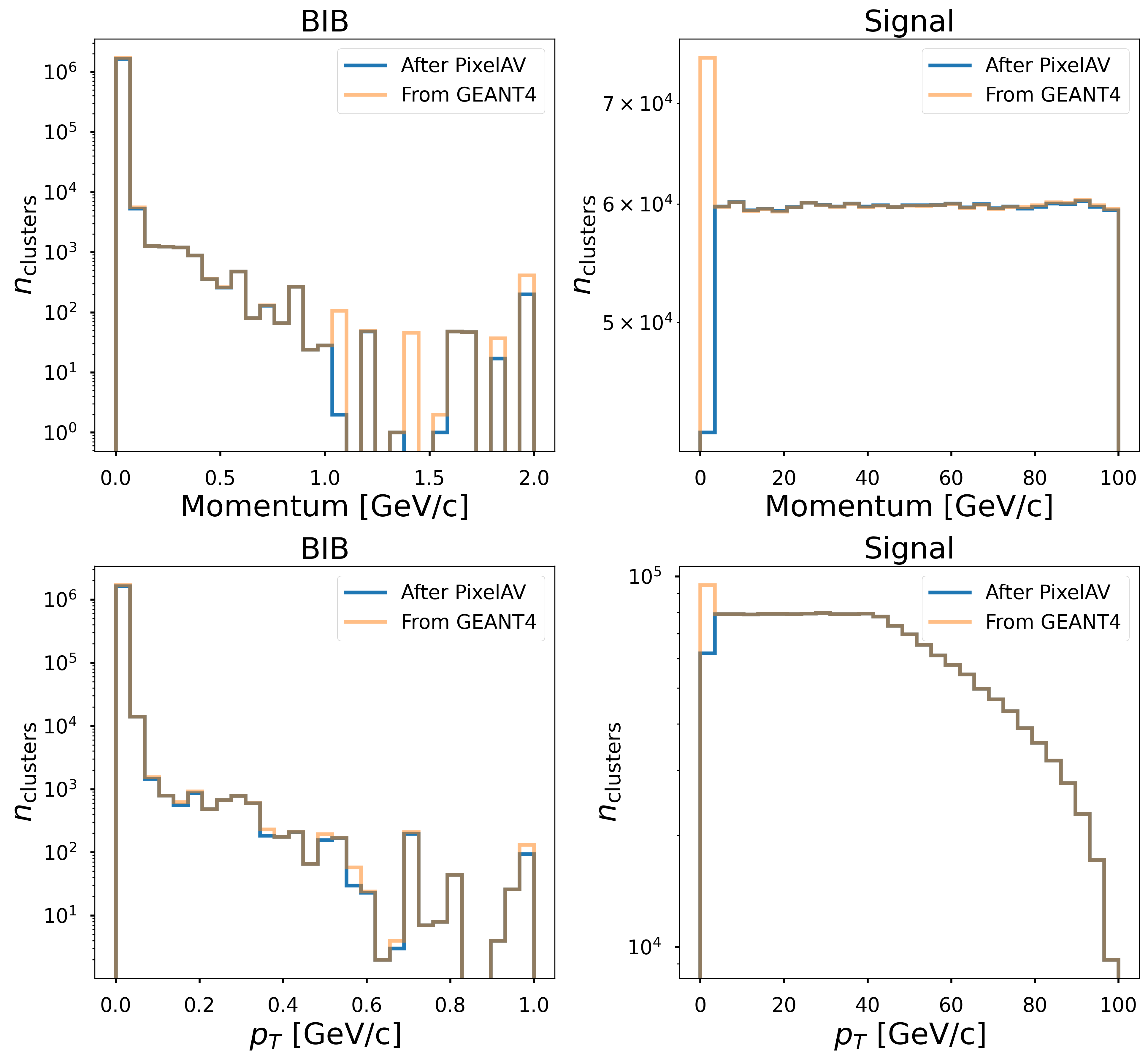}
    \caption{Distributions of the hit variables from \texttt{GEANT4} and the corresponding distributions of the same variables that are present in the clusters produced by \texttt{PixelAV}.}
    \label{fig:tracklistParquetMomenta}
\end{figure}

\begin{figure}
    \centering
    \includegraphics[width=\linewidth]{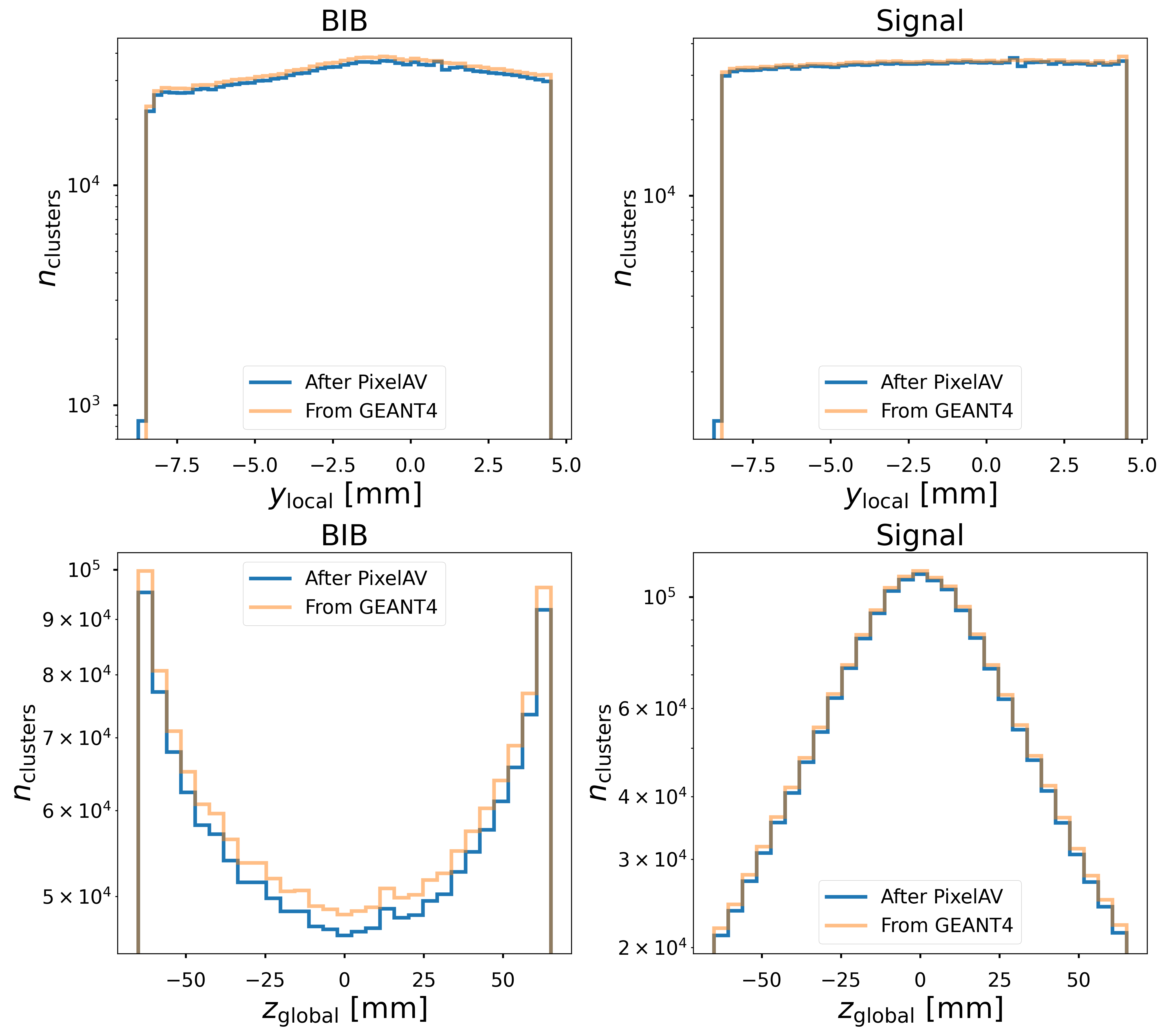}
    \caption{Distributions of the hit variables from \texttt{GEANT4} and the corresponding distributions of the same variables that are present in the clusters produced by \texttt{PixelAV}.}
    \label{fig:tracklistParquetCoord}
\end{figure}

\begin{figure}
    \centering
    \includegraphics[width=\linewidth]{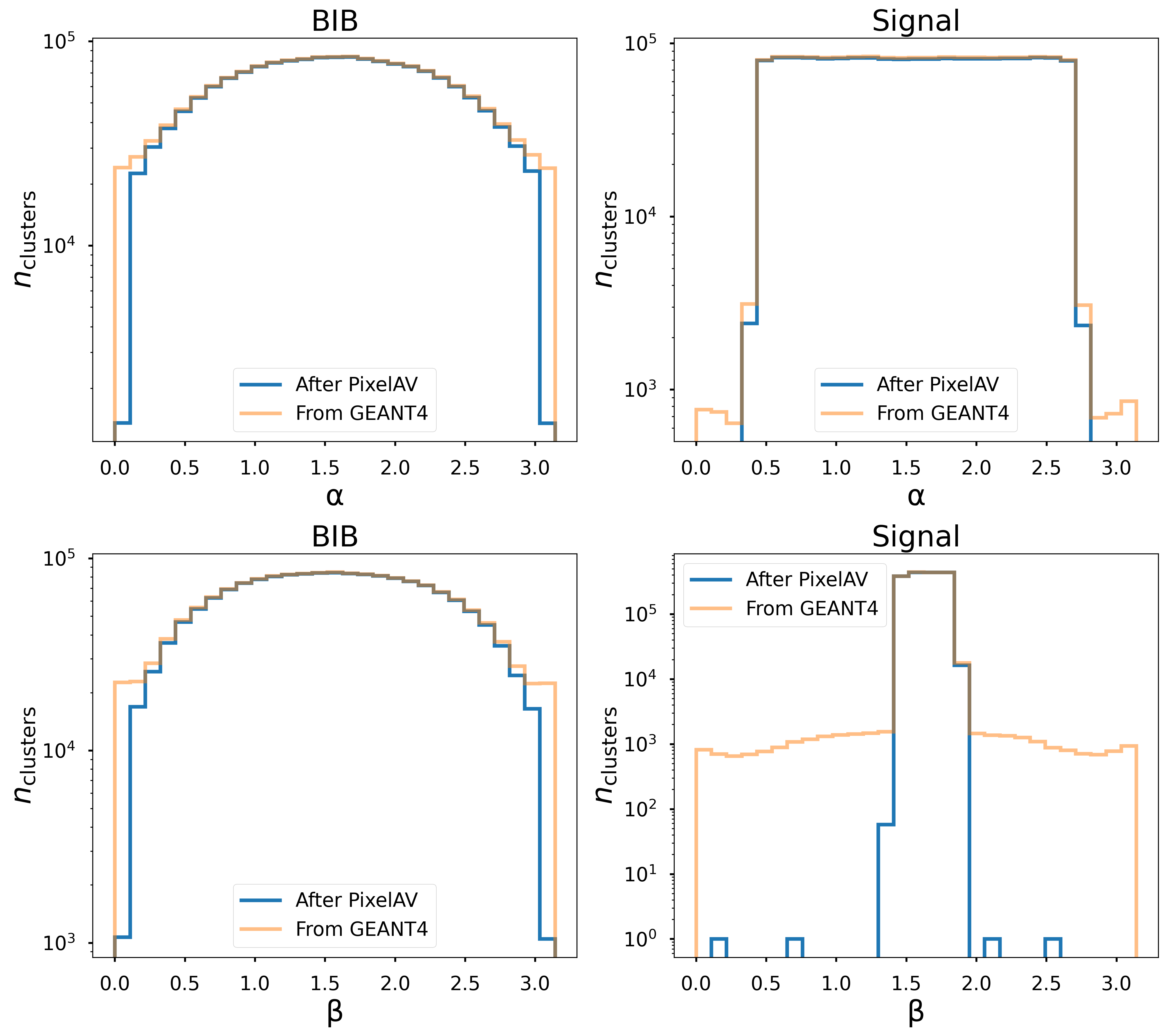}
    \caption{Distributions of the hit variables from \texttt{GEANT4} and the corresponding distributions of the same variables that are present in the clusters produced by \texttt{PixelAV}.}
    \label{fig:tracklistParquetAngs}
\end{figure}

Figures~\ref{fig:tracklistParquetMomenta}, \ref{fig:tracklistParquetCoord}, and \ref{fig:tracklistParquetAngs} show the distributions of the hit variables from \texttt{GEANT4} and the corresponding distributions of the same variables that are present in the clusters produced by \texttt{PixelAV}. 


\FloatBarrier
\section{Appendix B: Expanded Machine Learning Results}
\label{app:mlresults}
All models were trained using an NVIDIA RTX277 A4000 GPU. We used the 2024.1 release of \texttt{Vitis HLS} and the 2026.2 release of  \texttt{Siemens Catapult  HLS}.
When implementing a neural network on FPGA, an implementation with nonprogrammable weights is often chosen because the whole firmware can be reprogrammed to update weights or the neural network architecture. However, on an ASIC the implementation is fixed, so programmable weights are necessary for reconfigurability. We present synthesis with and without programmable weights, at various target clocks, in Table \ref{table:bkrejAndHW_expandedSynthesis}. All these syntheses were performed with the ``Resource" \texttt{hls4ml} strategy. We also present syntheses performed with the ``Latency" hls4ml strategy for nonprogrammable weights, which would also be interesting for an FPGA implementation in Table \ref{table:bkrejAndHW_Latency}.

An expanded overview of the hyperparameter scan is presented in Table \ref{tab:hyperparameter-search}.

\begin{table}[h]
\centering
\begin{tabular}{l|l|c|cc|c}
\hline
\multicolumn{2}{l|}{\textbf{Model Architecture}} & \textbf{Model 1} & \multicolumn{2}{c|}{\textbf{Model 2}} & \textbf{Model 3} \\
\hline
\multicolumn{2}{l|}{\textbf{Weight Quantization}} & 8-bit & 10-bit & 10-bit & 10-bit \\
\hline
\multicolumn{2}{l|}{\textbf{Number of Parameters}} & 265 & 711 & 2,637 & 14,241 \\
\hline
\multirow{3}{*}{\begin{tabular}{@{}l@{}}\textbf{Cluster}\\\textbf{Rejection} \\\textbf{Rate}\end{tabular}} 
& \bibrej{95}   & 86.9 \% & 88.9 \% & 91.8 \% & 93.8 \% \\ 
& \bibrej{98}  & 84.7 \% & 86.8 \% & 89.5 \% & 92.1 \% \\ 
& \bibrej{99}  & 82.0 \% & 84.6 \% & 87.1 \% & 90.5 \% \\
\hline
\multirow{2}{*}{\begin{tabular}{@{}l@{}}\textbf{Data}\\\textbf{Reduction} \end{tabular}} 
 & \bibred{99}  & 88.4 \% & 90.4 \% & 91.8 \% & 94.4 \% \\
 & \bibred{99} + $[3\sigma_t, 5\sigma_t]$  & 98.5 \% & 98.7 \% & 98.9 \% & 99.3 \% \\
\hline

\hline
\multirow{6}{*}{\begin{tabular}{@{}l@{}}\textbf{FPGA}\\\textbf{Resource} \\\textbf{Usage}\\\textbf{Nonprogrammable}\\\textbf{Weights}\\ \textbf{(v-synth)}\end{tabular}} 
    & LUT      & 9,132   & 34,255  & 111,947 & 89,400  \\
    & FF       & 4,266   & 11,965  & 28,375  & 69,951  \\
    & DSP      & 0       & 0       & 0       & 0      \\
    & BRAM     & 0       & 0       & 0       & 75 \\
    & Clock Period [ns] & 10.271 &10.324 &10.499 & 10.392 \\
    & Target Clock [ns] &3 & 4   & 5      & 12  \\ 
    & Latency [$\mu$s] & 0.240 & 0.228   & 0.264    & 87.780 \\
    & Initiation Interval & 16 & 16  & 12       & 6553 \\
\hline
\multirow{6}{*}{\begin{tabular}{@{}l@{}}\textbf{FPGA}\\\textbf{Resource} \\\textbf{Usage}\\\textbf{Programmable}\\\textbf{Weights} \\\textbf{(20 ns clock)}\\ \textbf{(v-synth)}\end{tabular}} 
    & LUT      & 5,835   & 10,314  & 18,529  & 89,735  \\
    & FF       & 5,457   & 11,361  & 20,234  & 84,787  \\
    & DSP      & 0       & 0       & 0       & 0      \\
    & BRAM     & 0       & 0       & 0       & 0 \\
    & Clock Period [ns] &17.049 &16.693 &16.317 & 17.335 \\
    & Target Clock [ns] &20 & 20   & 20      & 20  \\ 
    & Latency** [$\mu$s]  & 2.720  & 7.160   & 25.880     & 256 \\
    & Initiation Interval** & 32 & 280  & 700       & 5734 \\
\hline
\multirow{6}{*}{\begin{tabular}{@{}l@{}}\textbf{FPGA}\\\textbf{Resource} \\\textbf{Usage}\\\textbf{Programmable}\\\textbf{Weights}\\\textbf{(10 ns clock)}\\ \textbf{(v-synth)}\end{tabular}} 
    & LUT      & 5,219   & 11,505  & 19,290 & 160,355  \\
    & FF       & 5,768   & 12,904  & 24,275  & 115,639  \\
    & DSP      & 0       & 0       & 0       & 0      \\
    & BRAM     & 0       & 0       & 0       & 0 \\
    & Clock Period [ns] &10.909 &11.388 &11.507 & 11.931 \\
    & Target Clock [ns] &10 & 10   & 10      & 10  \\ 
    & Latency** [$\mu$s]  & 1.582  & 4.157  & 14.971     & 166 \\
    & Initiation Interval** & 32 & 280  & 700       & 6826 \\
\hline
\multirow{6}{*}{\begin{tabular}{@{}l@{}}\textbf{FPGA}\\\textbf{Resource} \\\textbf{Usage}\\\textbf{Programmable}\\\textbf{Weights}\\\textbf{(12 ns clock)}\\ \textbf{(v-synth)}\end{tabular}} 
    & LUT      & 5,588   & 11,465  & 20,026 & 145,551  \\
    & FF       & 5,413   & 12,337  & 23,381  & 104,176  \\
    & DSP      & 0       & 0       & 0       & 0      \\
    & BRAM     & 0       & 0       & 0       & 0 \\
    & Clock Period [ns] &10.909 &11.388 &11.507 & 11.931 \\
    & Target Clock [ns] &12 & 12   & 12      & 12  \\ 
    & Latency** [$\mu$s]   & 1.692  &  4.344  & 15.576     & 164 \\
    & Initiation Interval** & 32 & 280  & 700       & 6553 \\
\hline
    
\end{tabular}
\caption{Quantization, background rejection, and FPGA v-synth resources of four representative configurations. Implementations are presented with programmable weights at varying target clock periods. 
*The maximum latency and initiation interval are presented because architectures do not have a deterministic pipeline. **Programmable weights require loading weights and biases with each inference, which unrealistically increases the latency and initiation interval.}
\label{table:bkrejAndHW_expandedSynthesis}
\end{table}

\begin{table}[h]
\centering
\begin{tabular}{l|l|c|cc|c}
\hline
\multicolumn{2}{l|}{\textbf{Model Architecture}} & \textbf{Model 1} & \multicolumn{2}{c|}{\textbf{Model 2}} & \textbf{Model 3} \\
\hline
\multicolumn{2}{l|}{\textbf{Weight Quantization}} & 8-bit & 10-bit & 10-bit & 10-bit \\
\hline
\multicolumn{2}{l|}{\textbf{Number of Parameters}} & 265 & 711 & 2,637 & 14,241 \\
\hline
\multirow{3}{*}{\begin{tabular}{@{}l@{}}\textbf{Cluster}\\\textbf{Rejection} \\\textbf{Rate}\end{tabular}} 
& \bibrej{95}   & 86.9 \% & 88.9 \% & 91.8 \% & 93.8 \% \\ 
& \bibrej{98}  & 84.7 \% & 86.8 \% & 89.5 \% & 92.1 \% \\ 
& \bibrej{99}  & 82.0 \% & 84.6 \% & 87.1 \% & 90.5 \% \\
\hline
\multirow{2}{*}{\begin{tabular}{@{}l@{}}\textbf{Data}\\\textbf{Reduction} \end{tabular}} 
 & \bibred{99}  & 88.4 \% & 90.4 \% & 91.8 \% & 94.4 \% \\
 & \bibred{99} + $[3\sigma_t, 5\sigma_t]$  & 98.5 \% & 98.7 \% & 98.9 \% & 99.3 \% \\
\hline

\hline
\multirow{6}{*}{\begin{tabular}{@{}l@{}}\textbf{FPGA}\\\textbf{Resource} \\\textbf{Usage}\\ \textbf{(v-synth)}\end{tabular}} 
    & LUT      & 6,505   & 31,842  & 111,948 & 492,029  \\
    & FF       & 2,208   & 13,495  & 47,202  & 201,191  \\
    & DSP      & 0       & 0       & 0       & 0      \\
    & BRAM     & 0       & 0       & 0       & 0 \\
    & Clock Period [ns] &9.275 &9.275 &9.275 & 8.745 \\
    & Target Clock [ns] &10 & 10   & 10      & 10  \\ 
    & Latency [ns]  & 160  & 160   & 200     & 3130* \\
    & Initiation Interval & 1 & 1  & 1       & 280* \\
\hline
    
\end{tabular}
\caption{Comparison including quantization, background rejection, and FPGA v-synth resources (nonprogrammable weights) of four representative configurations. This implementation was generated with the ``Latency'' Vitis strategy rather than the ``Resource'' strategy which is used for all other implementations presented. 
*The Model 3 architecture implementation does not have a deterministic pipeline, so the maximum latency and initiation interval are presented.}
\label{table:bkrejAndHW_Latency}
\end{table}


\begin{table}[htbp]
    \centering
    \small
    \renewcommand{\arraystretch}{1.25}
    \setlength{\tabcolsep}{6pt}

    \begin{tabularx}{\textwidth}{
        |>{\centering\arraybackslash}p{2.0cm}
        |>{\raggedright\arraybackslash}X
        |>{\centering\arraybackslash}p{3.2cm}|
    }
        \hline
        \textbf{Model}
        & \textbf{Hyperparameter or setting}
        & \textbf{Values} \\
        \hline

        \multirow{3}{*}{\textbf{1}}
        & Number of hidden dense layers
        & \(2,3,4,5\) \\
        \cline{2-3}

        & Neurons per hidden layer
        & \(2\)--\(11\) \\
        \cline{2-3}

        & Learning rate
        & \(10^{-3}\) \textit{(fixed)} \\
        \hline

        \multirow{6}{*}{\textbf{2}}
        & Profile-branch width
          (\texttt{x-profile}, \texttt{y-profile}, \texttt{y-local})
        & \(8,16,\ldots,128\) \\
        \cline{2-3}

        & Position-branch width
          (\texttt{n-Module}, \texttt{x-local})
        & \(2,4,\ldots,12\) \\
        \cline{2-3}

        & First post-merge width ratio
        & \(0.2,0.3,\ldots,0.7\) \\
        \cline{2-3}

        & Second post-merge width ratio
        & \(0.2,0.3,\ldots,0.7\) \\
        \cline{2-3}

        & Learning rate
        & \(10^{-4}\)--\(10^{-2}\) \\
        \cline{2-3}

        & Dropout rate
        & \(0.08\) \textit{(fixed)} \\
        \hline

        \multirow{7}{*}{\textbf{3}}
        & Number of convolutional filters
        & \(2,4,6,8,10\) \\
        \cline{2-3}

        & Convolutional kernel size
        & \(3\times3\) \textit{(fixed)} \\
        \cline{2-3}

        & Scalar-branch width
          (\texttt{n-Module}, \texttt{x-local}, \texttt{y-local})
        & \(8,16,24,32\) \\
        \cline{2-3}

        & First post-merge dense width
        & \(8,24,\ldots,120\) \\
        \cline{2-3}

        & Second post-merge width ratio
        & \(0.2,0.3,\ldots,0.8\) \\
        \cline{2-3}

        & Learning rate
        & \(10^{-4}\)--\(10^{-2}\) \\
        \cline{2-3}

        & Dropout rate
        & \(0.08\) \textit{(fixed)} \\
        \hline

        \multirow{4}{2.0cm}{\centering\textbf{All quantized models}}
        & Weight and bias precision, \(b_{\mathrm{w}}\)
        & \(3,4,6,8,10\) bits \\
        \cline{2-3}

        & Input precision
        & \(b_{\mathrm{w}}+2\) bits \\
        \cline{2-3}

        & Hidden-layer activation precision
        & 8-bit quantized ReLU \\
        \cline{2-3}

        & Output activation
        & 8-bit quantized sigmoid \\
        \hline
    \end{tabularx}

    \caption{
        Hyperparameter search spaces and fixed settings for Models~1--3.
        Listed sequences indicate discrete tuner values. For quantized models,
        \(b_{\mathrm{w}}\) denotes the weight and bias bit width. Inputs were
        represented using \(b_{\mathrm{w}}+2\) bits, while hidden-layer
        activations and the final sigmoid output used eight bits.
    }
    \label{tab:hyperparameter-search}
\end{table}

\end{document}